\documentclass{aa}  

\usepackage{graphicx}
\usepackage{txfonts}
\begin{document}

\title{Long-term spot monitoring of the young solar analogue \object{V889 Her} 
\thanks{Based on observations made with the Nordic Optical Telescope, operated on the island of La Palma jointly by Denmark, Finland, Iceland,
Norway, and Sweden, in the Spanish Observatorio del Roque de los
Muchachos of the Instituto de Astrofisica de Canarias. Based on observations 
made with the HARPSpol instrument on the ESO 3.6 m telescope at La Silla 
(Chile), under the programme ID 091.D-0836.}\thanks{Photometric data is only available in electronic form at the CDS via anonymous ftp to cdsarc.u-strasbg.fr (130.79.128.5) or via http://cdsweb.u-strasbg.fr/cgi-bin/qcat?J/A+A/}}

   \author{T. Willamo
          \inst{1}
          \and
          T. Hackman\inst{1}
          \and
          J. J. Lehtinen\inst{2}
          \and
          M. J. Käpylä\inst{2}$^,$ \inst{3}
          \and
          I. Ilyin\inst{4}
          \and
          G. W. Henry\inst{5}
          \and
          L. Jetsu\inst{1}
          \and
          O. Kochukhov\inst{6}
          \and
          N. Piskunov\inst{6}
          }

   \institute{Department of Physics, P.O. Box 64, FI-00014 University of Helsinki, Finland \\
              \email{teemu.willamo@helsinki.fi}
             \and
             Max Planck Institute for Solar System Research, Justus-von-Liebig-Weg 3, D-37077 Göttingen, Germany
             \and
             ReSoLVE Centre of Excellence, Department of Computer Science, Aalto University, PO Box 15400, FI-00076 Aalto, Finland
             \and
             Leibniz-Institute for Astrophysics Potsdam, An der Sternwarte 16, 14482 Potsdam, Germany
             \and
             Center of Excellence in Information Systems, Tennessee State University, Nashville, TN
             37209, USA
             \and
             Department of Physics and Astronomy, Uppsala University, Box 516, 751 20 Uppsala, Sweden
             }

   \date{Received 2 November 2018 / Accepted 12 January 2019}

 
  \abstract
   {Starspots are important manifestations of stellar magnetic activity. By studying their behaviour in young solar analogues, we can unravel the properties of their magnetic cycles. This gives crucial information of the underlying dynamo process. Comparisons with the solar cycle enable us to infer knowledge about how the solar dynamo has evolved during the Sun's lifetime.}
   {Here we study the correlation between photometric brightness variations, spottedness, and mean temperature in V889 Her, a young solar analogue. Our data covers 18 years of spectroscopic and 25 years of photometric observations.}
   {We use Doppler imaging to derive temperature maps from high-resolution spectra. We use the Continuous Period Search method to retrieve mean V-magnitudes from photometric data.}
{Our Doppler imaging maps show a persistent polar spot structure varying in 
  strength. This structure is centred  
slightly off the rotational pole. The
mean temperature derived from the maps shows an overall decreasing trend, as does the photometric mean brightness, until it reaches its minimum around 2017. The filling factor of cool spots, however, shows only a weak tendency to anti-correlate with the decreasing mean brightness.}
{We interpret V889 Her to have entered into a grand maximum in its activity. The clear relation between the mean temperature of the Doppler imaging 
surface maps and the mean magnitude supports the reliability of the Doppler
images. The lack of correlation between the mean magnitude and the spottedness
may indicate that bright features in the Doppler images are real.}


   \maketitle
%

\section{Introduction}

The single, late-type star \object{V889 Her} (or \object{HD 171488})
is a young solar analogue of spectral class G2 V, classified as a BY Draconis variable \citep{73rd}, with an estimated age of 30-50 Myr \citep{strassmeier03}. Young, solar-type stars are important for the study of stellar activity and evolution since they provide us a look into what the Sun might have been like in the past. In this early evolutionary stage the stars are much more magnetically active than the Sun today. 
An approximately 10-year activity cycle has been detected from photometry of V889 Her, with a current trend of decreasing brightness \citep{lehtinen16}. Many Doppler images of it have been published, with a large  high-latitude spot visible in all of them \citep[e.g.][]{strassmeier03,jarvinen08,frasca10}.

\cite{marsden06} also studied the surface magnetic field of V889 Her using Zeeman-Doppler imaging (ZDI) with Stokes I\&V spectra. They found no latitude dependence of the radial field, as opposed to the azimuthal field forming ring-like structures around the pole. A similar result was reported in the ZDI study by \cite{jeffers08}. Neither of these studies revealed any clear connection between the radial magnetic field and cool spots.

Due to its young age, V889 Her is still a very rapid rotator, with a rotational period $P_{\mathrm{rot}} \approx 1.33$ days 
\citep[e.g.][]{jarvinen08}. The crucial parameters of stellar dynamos that drive the magnetic activity are rotation and rotationally affected turbulent convection \cite[see e.g. the review of][]{brandenburg2005}. The dynamo activity and the resulting magnetic activity  thus increase as a function of rotation, at least to a certain limiting value \cite[e.g.][]{wright2011, wright2016}. In comparison to the Sun, the observations indicate that these stars exhibit spot structures that have filling factors much larger than the Sun. The resolution of the observations is not yet accurate enough to reveal whether these large spotted areas are individual spots or consist of groups of smaller spots. 
On rapid rotators, spots are commonly observed at high latitudes. This has been confirmed even with interferometric observations on $\zeta$ And \citep{roettenbacher}.

In this paper we present both spectroscopic and photometric observations of V889 Her, ranging for 18 and 25 years, respectively. 
From the spectroscopic observations we 
calculate 
surface temperature maps of the star, using the Doppler imaging method, and study the evolution of the mean temperature, and the spot coverage and distribution. From the photometry we follow the long-term brightness changes in the star, 
thus 
tracing 
its activity cycles.

\section{Observations}

\subsection{Spectroscopy}

Spectroscopic observations of V889 Her have been gathered using the SOFIN and FIES instruments at the Nordic Optical Telescope (NOT) at La Palma and HARPS at the ESO 3.6m telescope at La Silla. We have analysed  19 sets of spectra, ranging from 1999 to 2017. The observations are listed in Table \ref{obs}. Some of the earlier sets have already been published by \cite{jarvinen08}. To calculate the rotational phases we used their ephemeris

\begin{equation}
\mathrm{HJD}_{0} = 2449950.550 + 1.33697 \times E,
\end{equation}

\noindent where HJD$_0$ corresponds to phase zero.

The FIES observations were reduced with the standard 
FIEStool pipeline \citep{FIEStool}, 
and the HARPS observations by the REDUCE package \citep{piskunov02}. The SOFIN observations were reduced with the new SDS tool, which is described in more detail in Section \ref{SOFINreduction}. The spectral resolutions of HARPS and FIES are 120 000 and 67 000, respectively.

\subsubsection{SOFIN observations}

Most of the spectroscopic observations were carried out with the SOFIN high-resolution polarimetric
spectrograph \citep{tuominen1999,ilyin2000} mounted at the NOT.  
We used the medium resolution 
optical camera with 1\,m focal length, which provides the resolving power of about
80\,000 with the entrance slit size of $0.5''$. 
The spectra were imaged with 
a Loral
$2048\times 2048\times 15\,\mu\mbox{m}$ CCD which provides a limited wavelength
coverage at the selected spectral range totaling 24 
orders
having about 45\,\AA\ in length at 5500\,\AA.

The data in this 
paper were obtained with different spectral set-ups, which is why different
spectral regions were used for different seasons. 
A typical exposure time for V889\, Her is about 40\, min,
achieving a signal-to-noise ratio of about 270 depending on the seeing and sky
conditions, as well as air mass.

\subsubsection{SOFIN data reduction} \label{SOFINreduction}

The SOFIN data reduction 
was 
done on a generic software platform written in C++ on
Linux called the Spectroscopic Data Systems (SDS). SDS can be used to analyse data from different
\'echelle spectrographs with the same tools. The image processing pipeline
specifically designed to handle SOFIN data calibration flow and image specific
content provides an automated process from the beginning to the end product without
human interaction and relies on 
statistical inference and robust
statistics. 
The software numerical toolkit and graphical interface is 
designed upon previous software 
\citep{ilyin2000}, and its complete description is in preparation 
(Ilyin 2019, in prep.).

The 
image processing 
includes 
the following steps:

\begin{itemize}

\item Bias subtraction 
using 
the CCD overscan and variance estimation of the pixel
intensity in the source images;

\item Master flat-field correction for the CCD spatial pixel-to-pixel noise. The
master flat is obtained as the 
normalised
sum of 100 flat-field spectra with the slit
decker opened enough to overlap spectral 
orders;

\item Scattered light subtraction is achieved by a 2D smoothing spline fit to
the gaps between spectral 
orders of the science frames.
The resistant statistics, which has a higher breakdown point than robust statistics with regard to how unaffected it is by outliers, is used to define the
gaps; 

\item The \'echelle orders are defined from the target image by a 3D fit of the
elongated Gaussian profiles along polynomial 
paths of the 
spectral orders;

\item The wavelength solution for the ThAr images uses a 3D Chebyshev fit of the spectral
line positions, 
and 
order 
number
to their wavelengths. It
also uses resistant statistics to find the initial wavelength in the
image. Typically, all available 1200 ThAr lines are used in the fit which
results in about 7\,m/s error of the fit in the image centre.  The ThAr image is
taken prior to the science exposure during observations to alleviate bending and
flexure of the Cassegrain mounted spectrograph;

\item The optimal extraction of the 
spectral orders and cosmic spikes
elimination in the target image is done by constructing a spatial profile from
the image which is resampled and smoothed with 2D spline into a regular
grid. The flux in each pixel is obtained with a weighted linear least-squares
fit of the spatial profile at a given pixel to the data. The cosmic spikes
elimination is done prior to the fit with the use of robust statistics;

\item The wavelength calibration of the 
orders in the target
image from the ThAr wavelength solution is followed by the heliocentric
correction of the wavelength scale. We use NOVAS\ 2 \citep{kaplan1989} to perform
this correction as implemented in SDS;

\item The extracted images are subjected to a 2D smoothing spline fit to the
continuum. This is achieved by taking an average image of all observations which
follows by its continuum normalisation with the aid of the resistant statistics
to exclude spectral lines from the fit. The ratio of each individual image and
the normalised average, which removes most of the spectral lines in the ratio,
is then used to fit a smoothing 2D robust spline which constitutes the 
intermediate 
continuum for the individual observation;

\item The final continuum normalisation is done during each iteration of the Doppler imaging inversion by comparing the observed spectrum to the modelled spectrum.

\end{itemize}

\subsubsection{Quality of the data sets}

To estimate the reliability of the temperature inversions produced from our observations, we have calculated the phase coverage $F$ of each set of spectra by assuming that one spectrum covers 10\% of one rotation, as was done by \cite{kochukhov13}. The phase coverage of our data sets vary from very good (the whole rotation is covered in Jul 2005, i.e. $F = 1.0$) to quite poor (in Oct 1999 $F = 0.325$); see Table \ref{obs} for a full list of the coverage.  The average signal-to-noise ratio ($S/N$) and the deviation in the final Doppler imaging inversions ($\sigma$) are also shown there as an indicator of the quality of each set of spectra.

\subsection{Photometry}

We analysed V-band photometry of V889 Her obtained between the years 1994 and 2018 with the Tennessee State University T3 0.4m Automatic Photoelectric Telescope (APT) at Fairborn Observatory in southern Arizona.

The observations of V889 Her were made differentially with respect to the comparison star HD 171286 and the check star HD 170829. The check minus comparison differential magnitudes show that both stars are constant to better than 0.005 mag from night to night and to 0.001 mag from year to year. The V magnitude of HD 171286 \citep[V=6.84;][]{oja87} was added to the variable minus comparison star differential magnitudes to derive the V magnitudes of V889 Her. See \cite{lehtinen16} for further information on the acquisition and reduction of V889 Her observations with the T3 APT.

\begin{table*}
\centering
\caption{Spectroscopic observations of V889 Her. Dates, HJD interval, number of
spectra ($N$), esimated phase coverage ($F$), mean signal-to-noise ratio 
($\langle S/N \rangle $), deviation between the Doppler imaging solution and the observations ($\sigma$) and instrument.}
\label{obs}
\begin{tabular}{c c c c c c c}
\hline\hline
Date & HJD-2400000 & $N$ & $F$ & $\langle S/N \rangle$ & $\sigma [\%]$ & Instrument\\
\hline
Jul -- Aug 1999 & 51383.482 -- 51394.511 & 11 & 0.596 & 252 & 0.455 & SOFIN\\
Sep 1999 & 51443.401 -- 51449.387 & 6 & 0.434 & 199 & 0.743 & SOFIN\\
Oct 1999 & 51471.359 -- 51475.337 & 4 & 0.325 & 285 & 0.400 & SOFIN\\
Jul -- Aug 2001 & 52120.452 -- 52131.470 & 12 & 0.595 & 234 & 0.457 & SOFIN\\
Aug 2002 & 52507.447 -- 52515.429 & 8 & 0.552 & 226 & 0.490 & SOFIN\\
Jul -- Aug 2004 & 53215.532 -- 53228.485 & 11 & 0.723 & 239 & 0.786 & SOFIN\\
Jul 2005 & 53567.406 -- 53577.562 & 16 & 1.0 & 250 & 0.511 & SOFIN\\
Aug -- Sep 2006 & 53979.443 -- 53991.449 & 6 & 0.476 & 225 & 0.651 & SOFIN\\
Jul 2007 & 54300.520 -- 54311.576 & 15 & 0.950 & 235 & 0.500 & SOFIN\\
Sep 2008 & 54717.368 -- 54723.486 & 8 & 0.703 & 228 & 0.680 & SOFIN\\
Aug -- Sep 2009 & 55069.520 -- 55081.520 & 14 & 0.606 & 196 & 0.760 & SOFIN\\
Jul 2010 & 55395.516 -- 55408.642 & 10 & 0.707 & 208 & 0.682 & SOFIN\\ 
Aug -- Sep 2012 & 56166.418 -- 56174.413 & 9 & 0.458 & 260 & 0.598 & SOFIN\\
Aug 2013 & 56521.424 -- 56525.446 & 5 & 0.408 & 147 & 0.744 & SOFIN\\
Sep 2013 & 56545.484 -- 56551.581 & 7 & 0.479 & 210 & 0.723 & HARPS \\
Aug 2014 & 56881.493 -- 56889.454 & 10 & 0.710 & 344 & 0.386 & FIES\\
Jul 2015 & 57204.511 -- 57213.563 & 10 & 0.657 & 299 & 0.482 & FIES\\
Jun 2016 & 57555.553 -- 57563.571 & 11 & 0.764 & 259 & 0.558 & FIES\\
Jun 2017 & 57908.615 -- 57918.531 & 11 & 0.496 & 116 & 0.882 & SOFIN\\
\hline
\end{tabular}
\end{table*}


\begin{table}
\centering
\caption{Stellar parameters of V889 Her.}
\label{param}
\begin{tabular}{c c c}
\hline\hline
Parameter & Value & Reference \\
\hline
Spectral class & G2 V & \cite{montes2001} \\
Age & 30-50 Myr & \cite{strassmeier03} \\
Period & 1.33697 d & \cite{jarvinen08} \\
$v\sin i$ & 38.5 km/s & This work \\
Inclination & $60^\circ$ & \cite{marsden06} \\
Microturbulence & 1.6 km/s & \cite{jarvinen08} \\
Macroturbulence & 3.0 km/s & \cite{strassmeier03}\\
\hline
\end{tabular}
\end{table}

\section{Doppler imaging}

Inhomogeneities in the surface temperature of a star, caused mainly by cool spots, leave traces in photospheric absorption lines, which can be tracked if the lines are sufficiently broadened by the star's rotation. These line profiles can be inverted to the surface temperature distribution which would cause the observed line profiles. For this purpose we use a modified version of the Doppler imaging code INVERS7, originally written by \cite{piskunov90}. 
For an optimal solution, INVERS7 uses Tikhonov regularisation to damp high spatial frequencies of temperature gradients. Recent improvements in the code include a molecular equilibrium solver \citep{piskunov2017}.

To reproduce the stellar spectra we use MARCS\footnote{http://marcs.astro.uu.se/} model atmospheres for temperatures in the range  3500-6500 K \citep{gustafsson08}, combined with absorption line parameters obtained from the Vienna Atomic Line Database\footnote{http://vald.astro.uu.se/} (VALD) \citep{piskunov95,kupka99}. In the inversion we have used 
spectral regions including
seven 
dominant 
absorption 
lines: the Fe I line at 6265.1319 Å, the V I line at 6266.3069 Å, the Fe I line at 6411.6476 Å, the Fe I line at 6430.8446 Å, the Fe II line at 6432.6757 Å, the Ca I line at 6439.0750 Å, and the Ni I line at 6643.6304 Å. Their line parameters are given in Table \ref{lin}. For the Fe II and V I lines and one of the Fe I lines we used the standard values for $\log(gf),$ as found  in the database, but for the two other Fe I lines and the Ca I and Ni I lines we modified the value slightly, as shown in Table 
\ref{lin}, in order to get consistent line strengths in the different regions. We chose to adjust the
$gf$-values, although the inconsistency could be caused by different effects related to difficulties in
modelling stellar line profiles using standard LTE-calculations. Adjusting the $gf$ -values is equivalent to changing the strength of individual lines. If the Doppler image is calculated using just one spectral line, the adjustment changes the mean temperature of the image, but not the spot configuration.
When using multiple lines, some of the modelled lines may be too strong while others are too weak compared to the observations. This will prevent reaching a satisfactory Doppler imaging solution, if no adjustments are done. Since the aim here is not to calculate
absolute stellar parameters or element abundances, adjusting the $gf$-values is acceptable.

In the SOFIN data the exact range of the available spectrum 
varies due to different instrument setups, 
so the same lines could not be used for all data sets. When available, we  used lines 3-6, as listed in Table \ref{lin};  
when these lines were not available, we  used 
other combinations. 
In Aug 2013 line 6 was not used, although available, due to obvious errors in some of the observed phases. 
The lines used in each set are listed along with the results in Table \ref{f-factors}. In the FIES and HARPS observations, the full visible spectrum is always available, so we could always use lines 3-6. Line 6 is the only one available in all sets for all instruments (although it was not used in Aug 2013), so we tried to account for the effects of using different line combinations by also repeating the inversions using only this Ca I line. 

In addition to the strongest lines, we included 91 atomic lines and 607 molecular lines (mainly TiO and CN) from four different wavelength regions (6260-6269 Å, 6405-6417 Å, 6425-6443 Å, and 6639-6648 Å) in the spectral line synthesis. They are not strong enough to be distinguished individually, but they can still affect the continuum level.

A telluric absorption line was cut away from the spectra around 6432-6433 Å. The exact location of the line varies between the spectra because they have been corrected for the Earth's motion to the rest frame of V889 Her. This overlaps heavily with the Fe II line at 6432.6757 Å, but the parts of the line that remain after the removal of the atmospheric line are still useful.

Important 
stellar 
parameters for the inversion process are listed in Table \ref{param}. These include the rotation period, $v\sin i$, the inclination, and micro- and macroturbulence. Similarly to \cite{jarvinen08}, we have used solar element abundances for V889 Her. As an initial guess for the temperature, we used 5800 K.

There have been several studies where significant differential rotation on V889 Her has been reported \citep{marsden06,jeffers08}, but also studies that yielded lower values for it \citep{huber09,jarvinen08} or no differential rotation at all \citep{kovari11}. Observations \citep{reiners03}, theory \citep{kitchatinov1999}, and global magnetoconvection simulations \citep{viviani2018} show that strong differential rotation should not be common in rapid rotators, such as V889 Her. Since the studies focusing on differential rotation yield very inconsistent results, we have excluded it from our study, assuming the star to rotate as a solid body. This is not necessarily true, but a moderate differential rotation should not significantly affect the general picture of spottedness, although it can affect the details of the spot configuration. Nevertheless, our data can be successfully reproduced without differential rotation.

\begin{table}
\centering
\caption{Parameters for the absorption lines used in the inversion.}
\label{lin}
\begin{tabular}{c c c c c}
\hline\hline
\# & Line & $\lambda (Å)$ & $\log(gf)$ & $\log(gf)_{\mathrm{standard}}$\\
\hline
1 & Fe I & 6265.1319 & -2.850 & -2.550\\
2 & V I & 6266.3069 & -2.290 & -2.290 \\
3 & Fe I & 6411.6476 & -0.620 & -0.695\\
4 & Fe I & 6430.8446 & -2.106 & -2.106 \\
5 & Fe II & 6432.6757 & -3.520 & -3.520 \\
6 & Ca I & 6439.0750 & 0.450 & 0.390 \\
7 & Ni I & 6643.6304 & -1.920 & -2.220 \\
\hline
\end{tabular}
\tablefoot{$\log(gf)$ is the adopted value for the line (modified from the standard value in some cases), while $\log(gf)_{\mathrm{standard}}$ is the standard value retrieved from VALD.}
\end{table}

\section{Time series analysis}

To analyse photometric data, we use the Continuous Period Search method (CPS) developed by \cite{CPS}. The method works by dividing the data into smaller 
sets and using a sliding window of constant time length. 
This enables the light curve parameters to change between each 
(usually overlapping) set. 
To each set, a function of the form

\begin{equation}
y(t_\mathrm{i})=M+\sum_{k=1}^K \Big(B_\mathrm{k}\cos(k2\pi ft_\mathrm{i})+C_\mathrm{k}\sin(k2\pi ft_\mathrm{i})\Big)
\end{equation}

\noindent is fitted. Here, $M$ is the mean magnitude of the set, $f = 1/P_\mathrm{rot}$ is the rotational frequency, and $t_\mathrm{i}$ the time of the data point. The best 
estimates 
for $M$, $f$, and the coefficients $B_\mathrm{k}$ and $C_\mathrm{k}$ are 
retrieved 
for each set. The order $K$ of the model is determined by using the Bayesian information criterion. An upper limit $K=2$ is used, so models with 0, 1, and 2 orders are tested. The length of one data set is defined by a maximum time span $\Delta t_\mathrm{max}$, such that all data points within the time $\Delta t_\mathrm{max}$ = 30 days of the first data point of the set are included. Only sets with at least 14 data points are analysed.

Here, we are primarily interested in the mean magnitude 
$M$. 
Periodic changes in $M$ are interpreted as signs of an activity cycle. By using the mean magnitude instead of individual measurements, the brightness variations caused by the star's rotation are diminished.

\cite{jetsu17} showed that the CPS method does not necessarily produce correct rotation period estimates if there are multiple interfering periodicities. However, this kind of interference will not affect the mean magnitudes we use here, while it would affect other light curve parameters, such as the amplitude. We have also extracted the phases of photometric minima, which could be affected by interfering periodicities.


\begin{figure*}[htb]
  \centering
  \begin{tabular}{cc}
    \includegraphics[width=7cm]{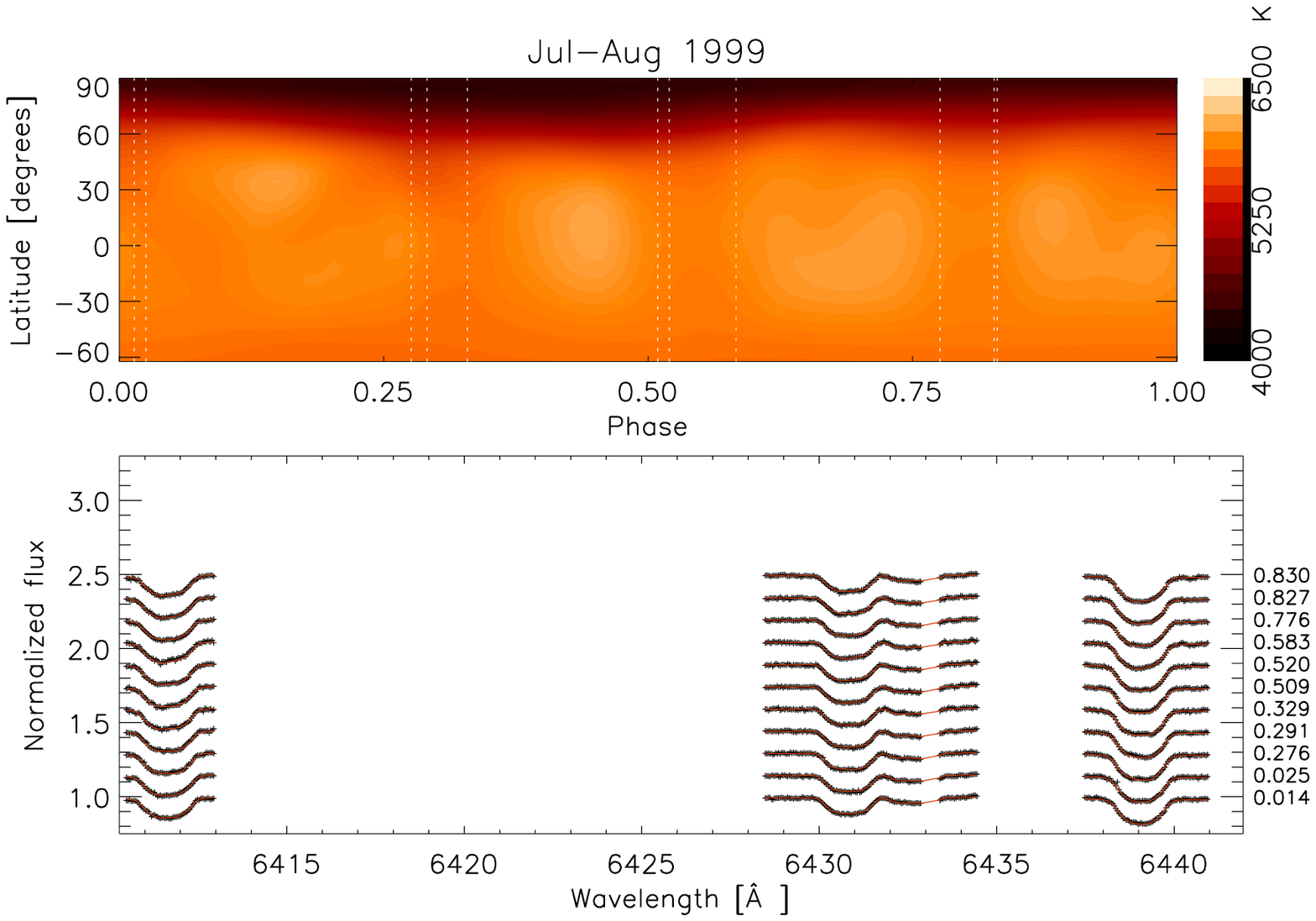} &
    \vspace{-0.3cm}
    \includegraphics[width=7cm]{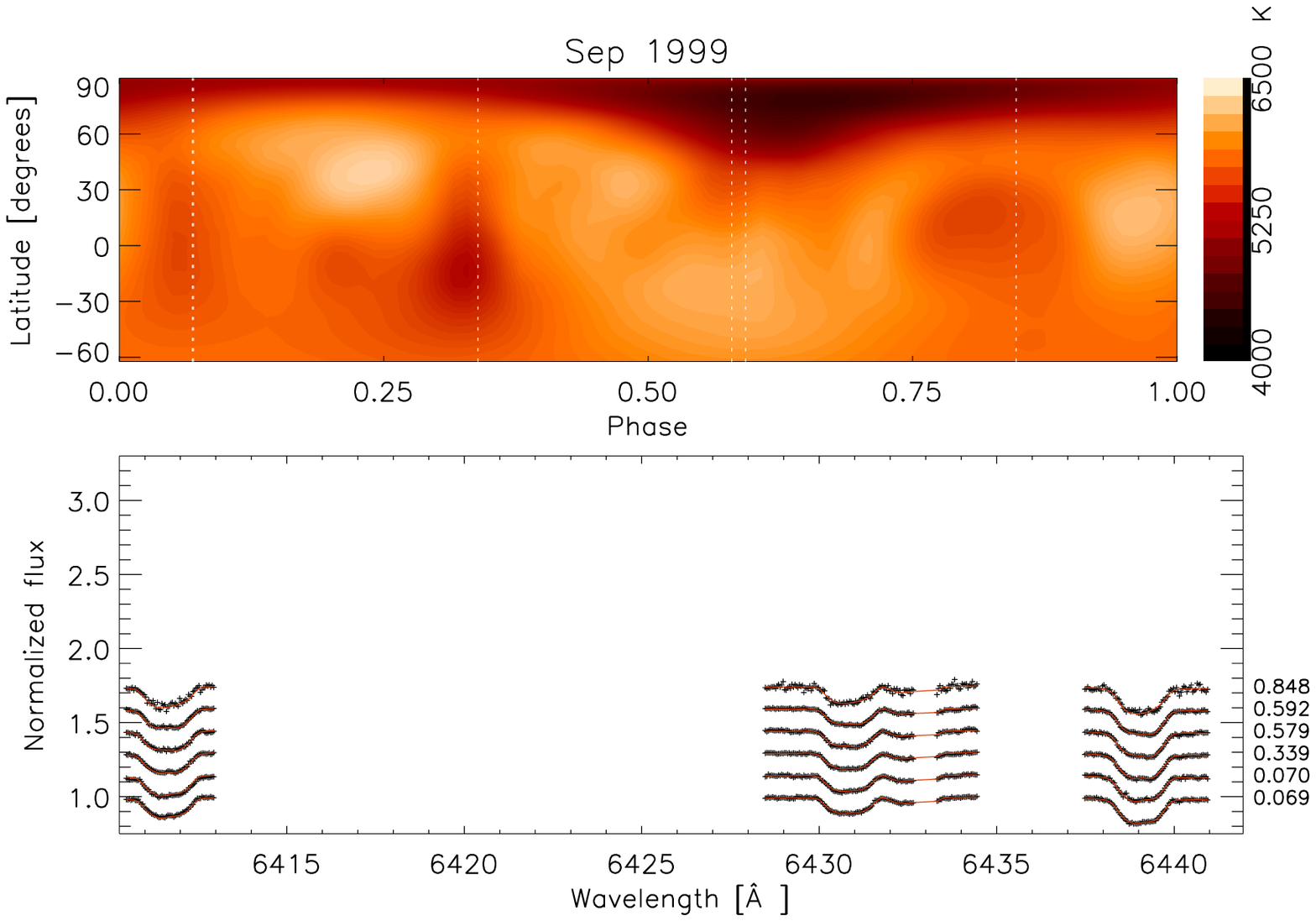} \\
    \vspace{-0.3cm}
    \includegraphics[width=7cm]{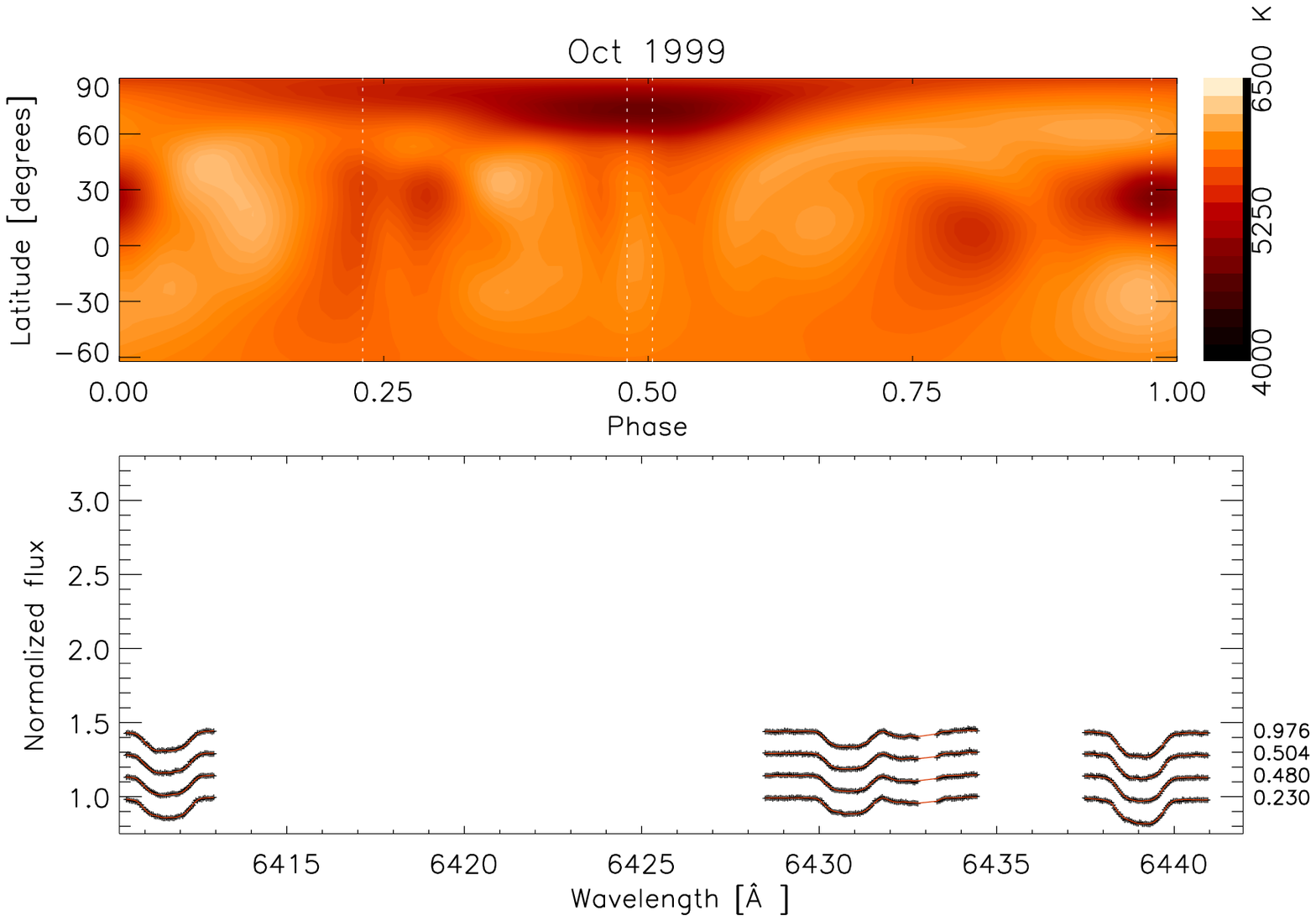} &
    \includegraphics[width=7cm]{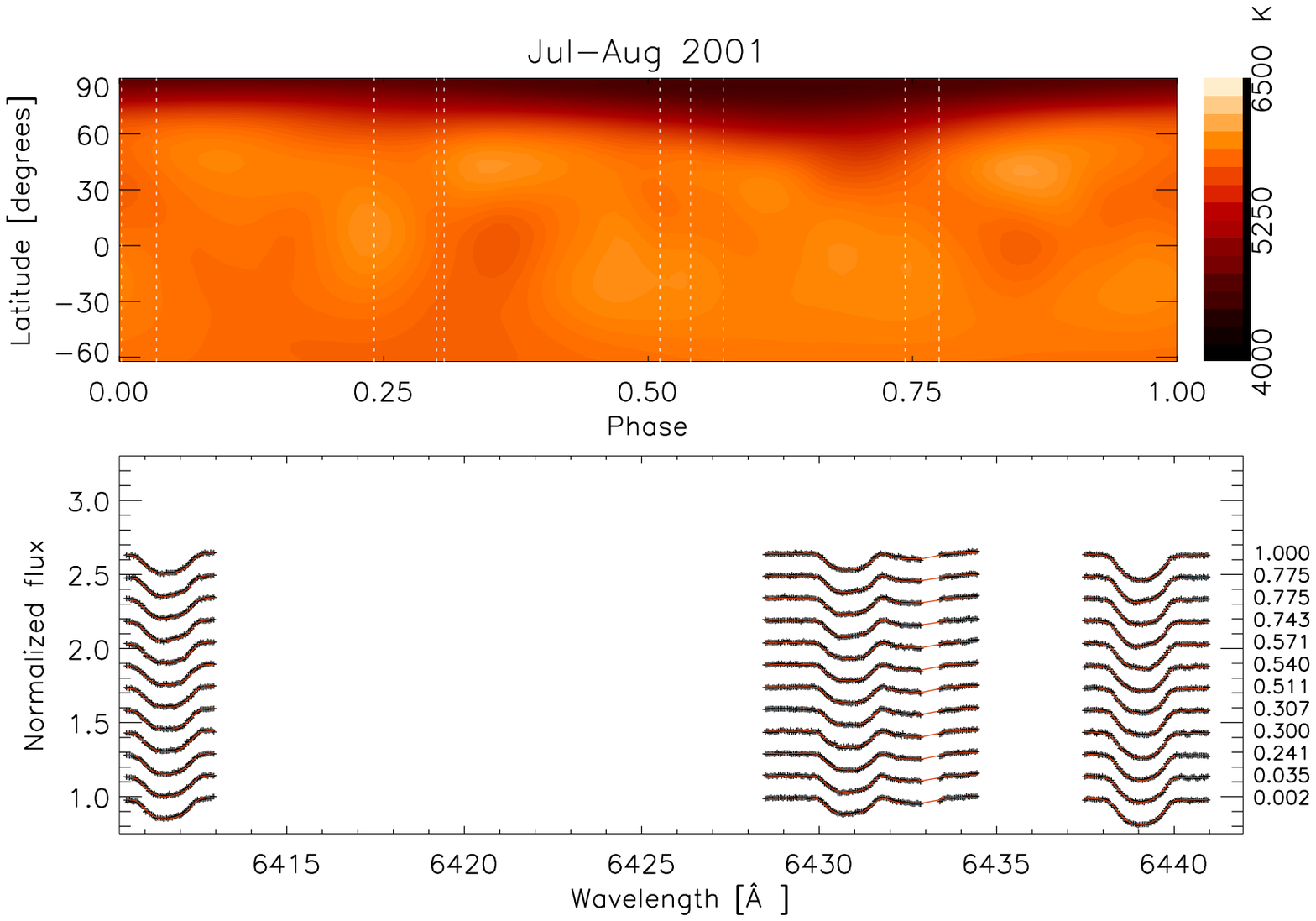} \\
    \vspace{-0.3cm}
    \includegraphics[width=7cm]{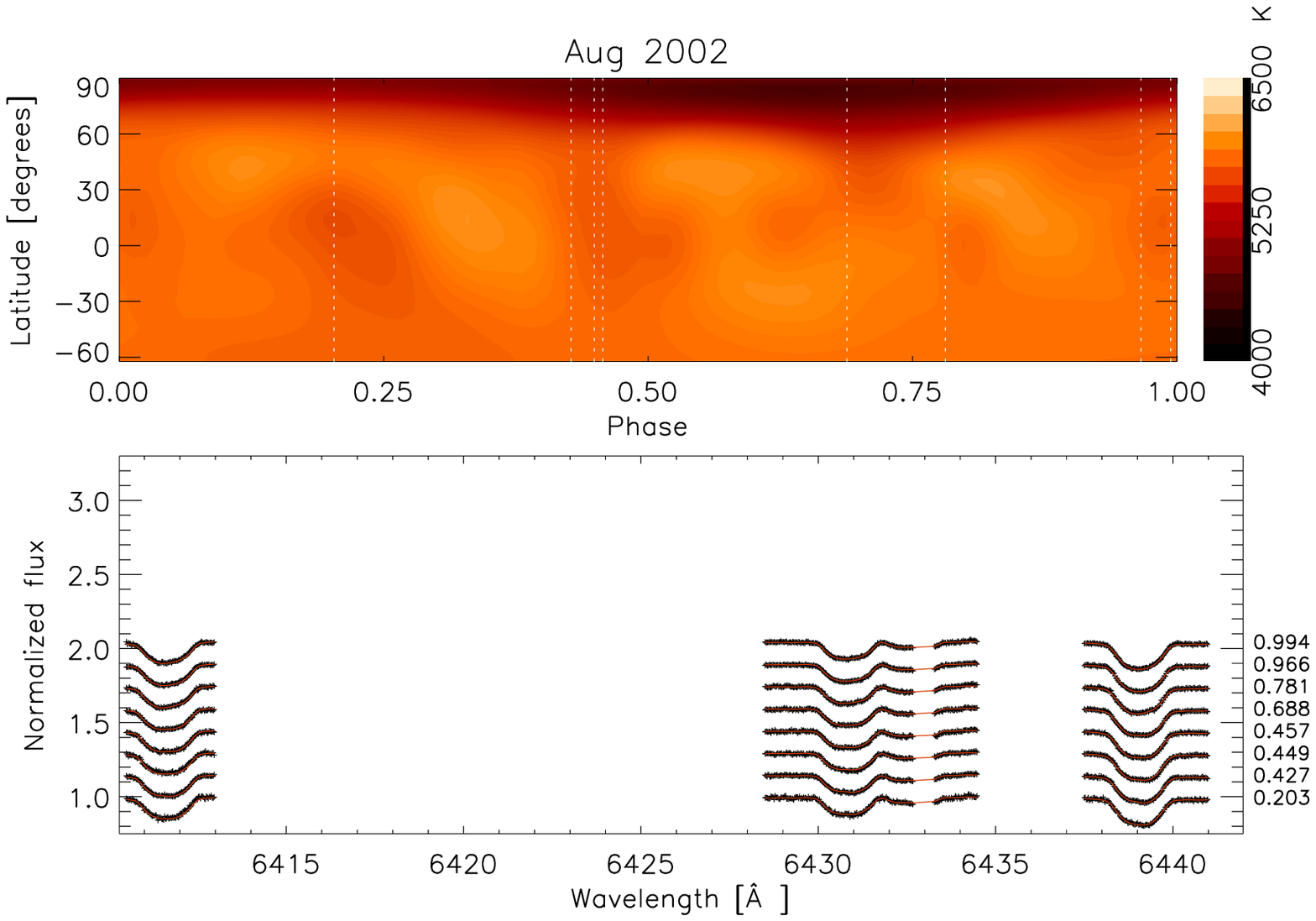} &
    \includegraphics[width=7cm]{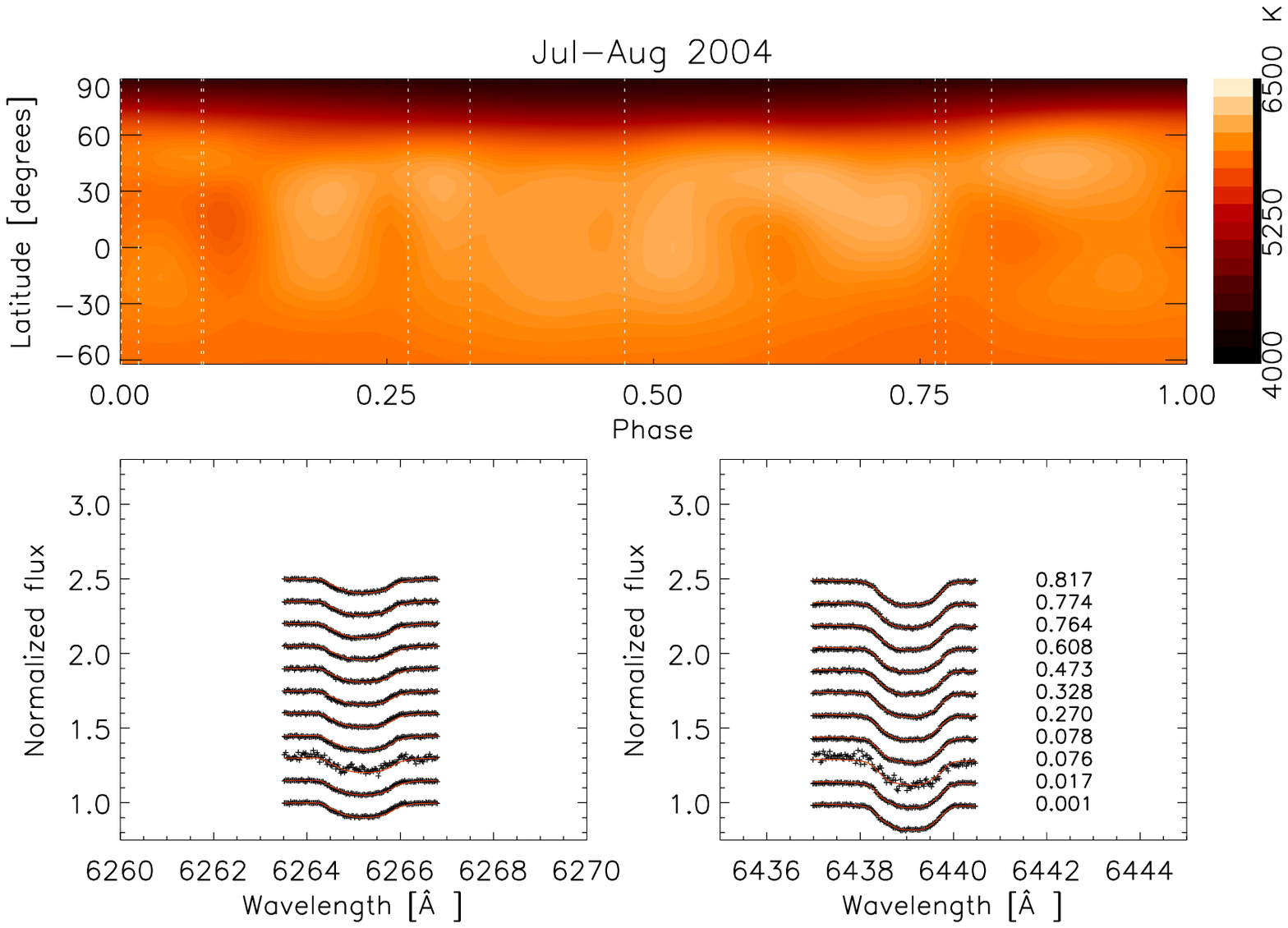} \\
    \vspace{-0.3cm}
    \includegraphics[width=7cm]{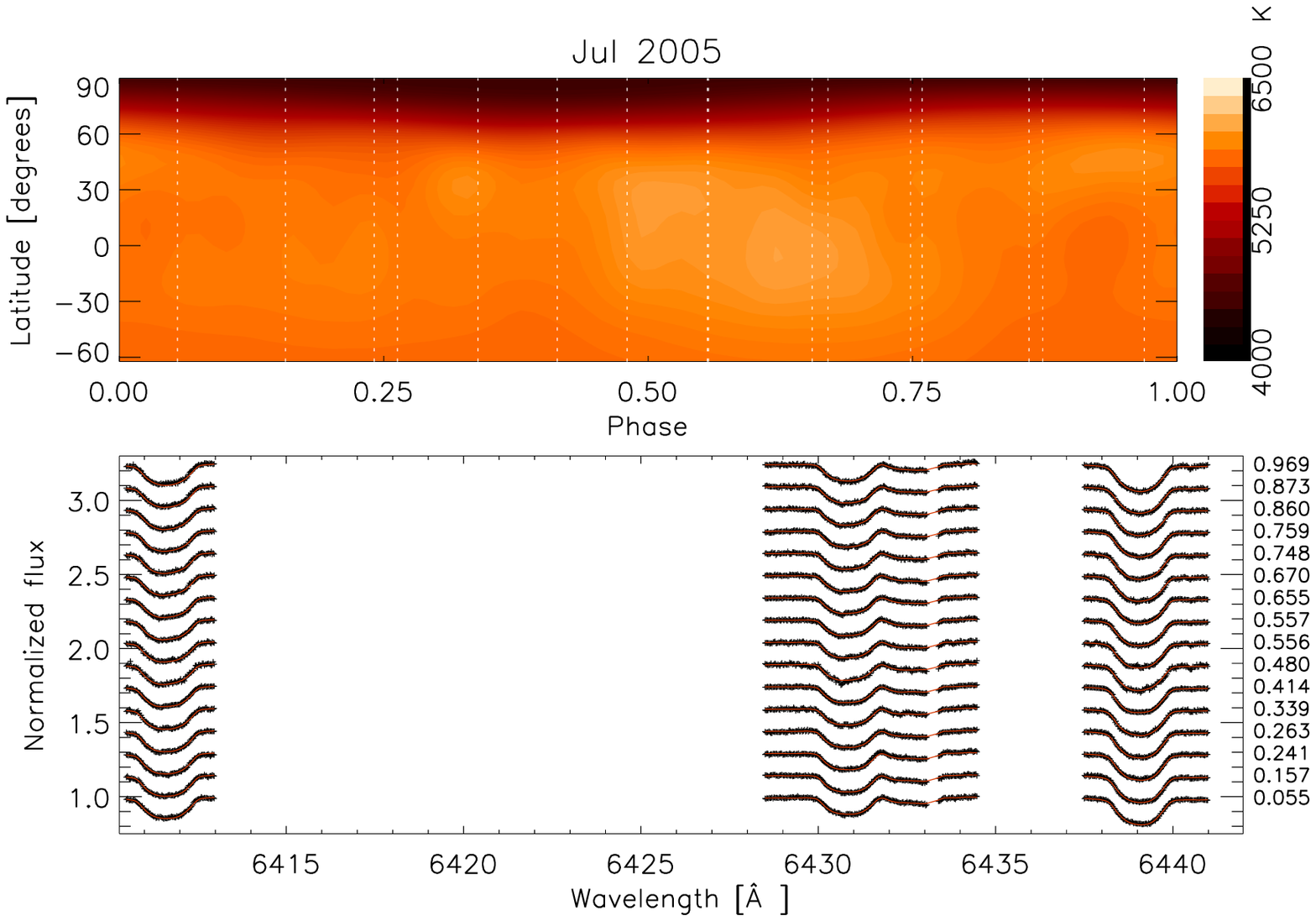} &
    \includegraphics[width=7cm]{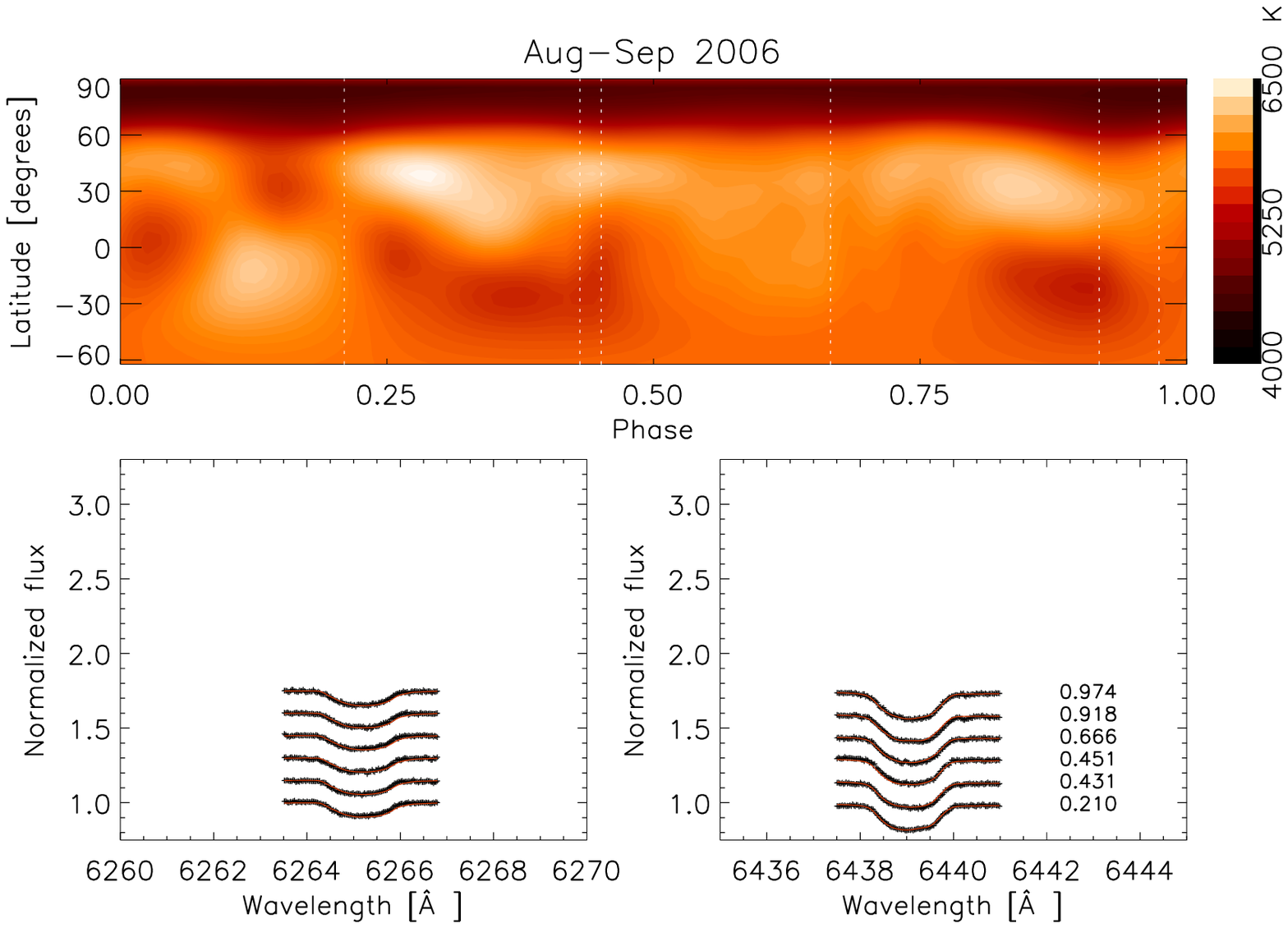} \\
    \vspace{-0.3cm}
    \includegraphics[width=7cm]{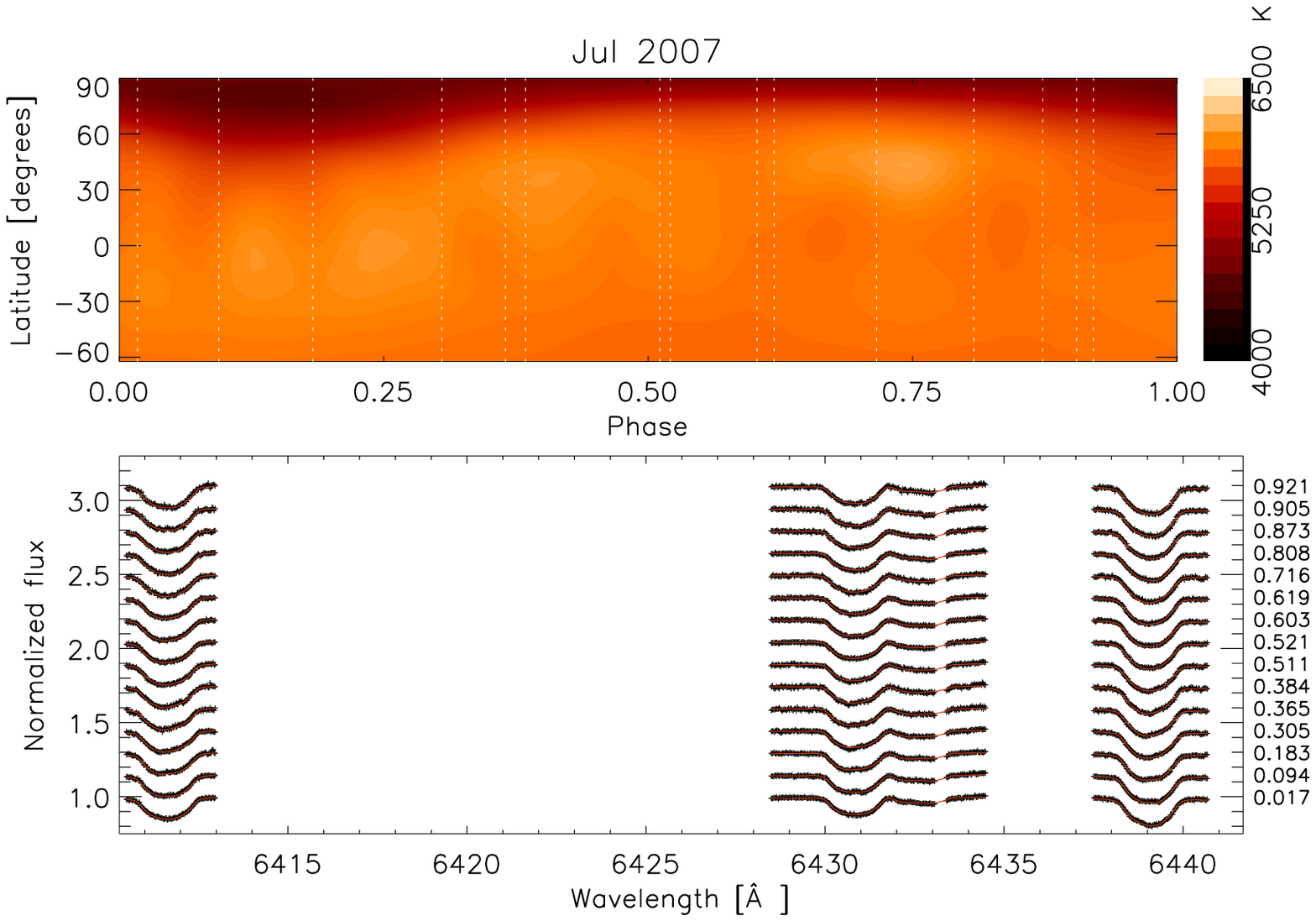} &
    \includegraphics[width=7cm]{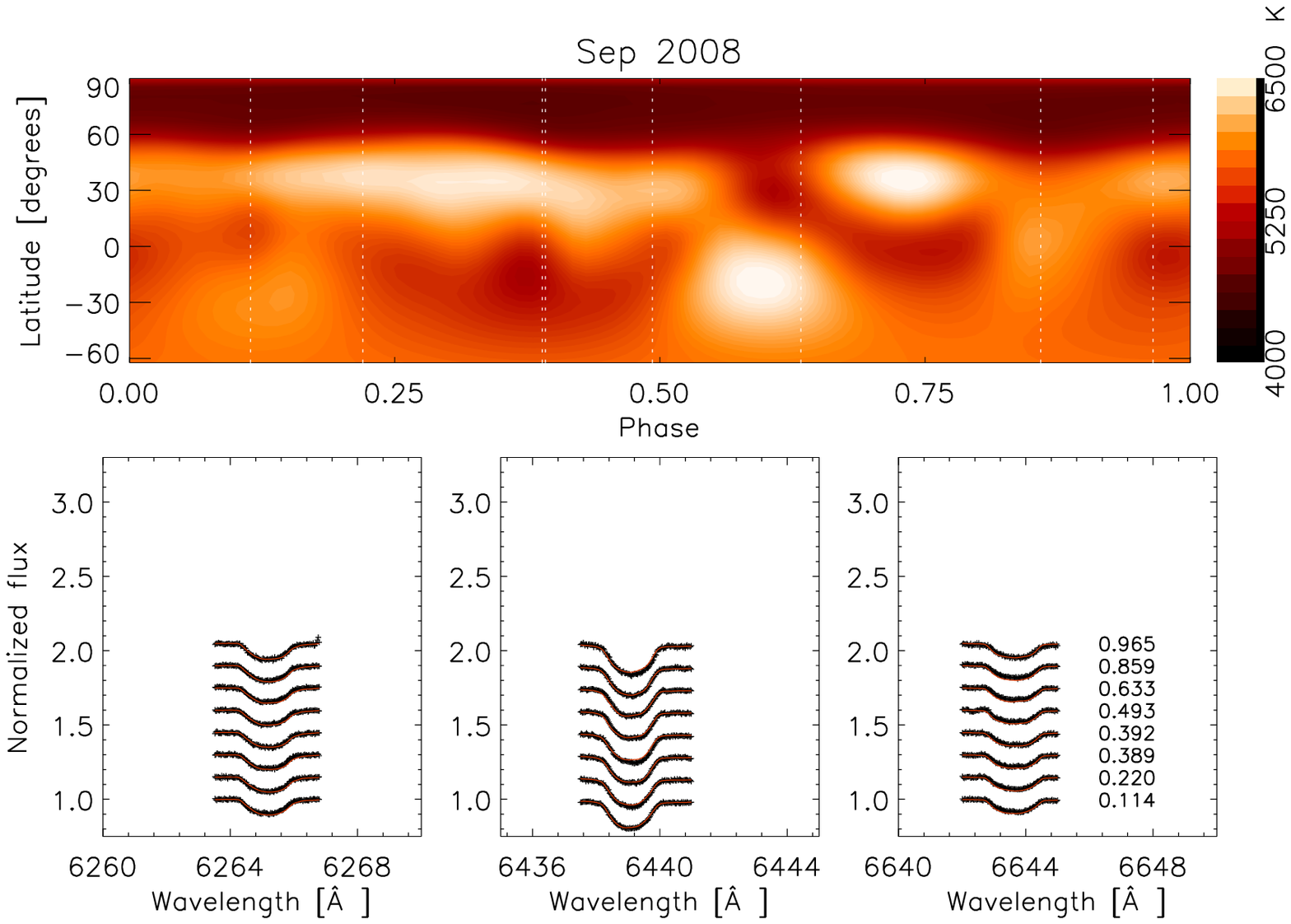} 
  \end{tabular}
  \caption{Doppler imaging maps of V889 Her for 1999-2008. Upper panels: Equirectangular projections of the surface temperature distribution. The longitude is given as the rotational phase. The vertical lines represent the phases of the individual spectra. Lower panels: Observed (black) and modeled (red) line profiles.}
  \label{DI_all}
\end{figure*}

\begin{figure*}[htb]
  \centering
  \begin{tabular}{cc}
    \includegraphics[width=7cm]{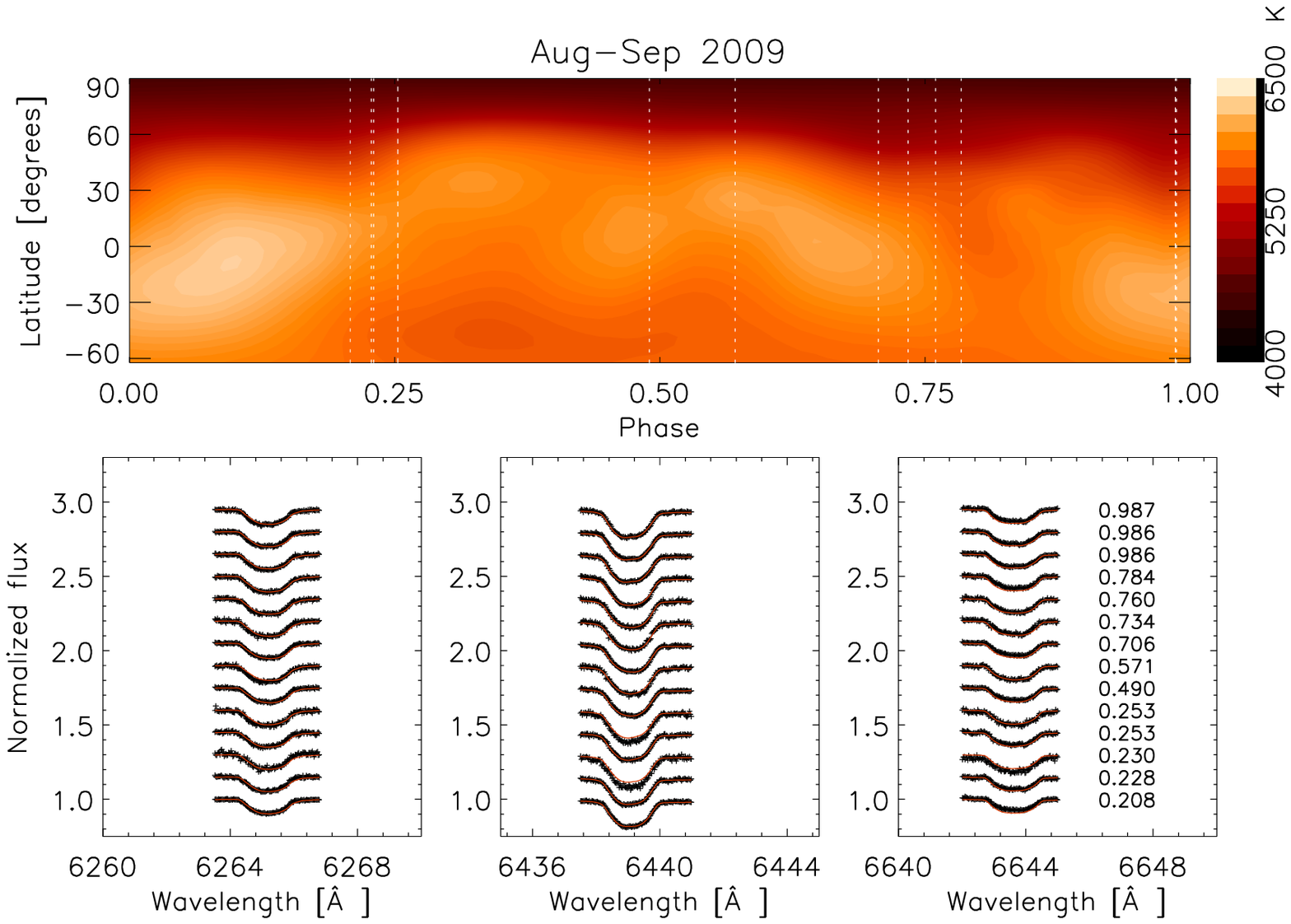} &
    \vspace{-0.3cm}
    \includegraphics[width=7cm]{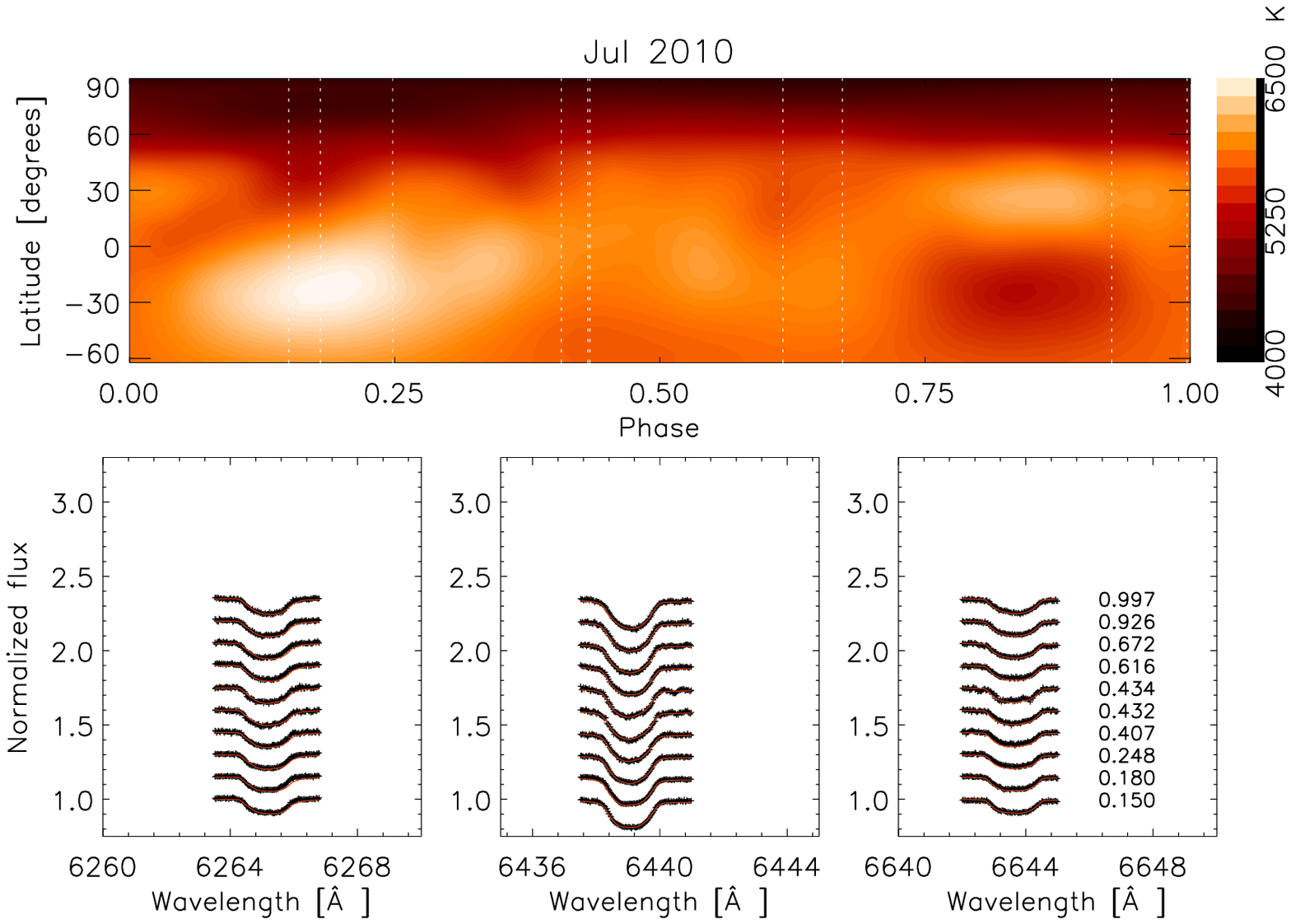} \\
    \vspace{-0.3cm}
    \includegraphics[width=7cm]{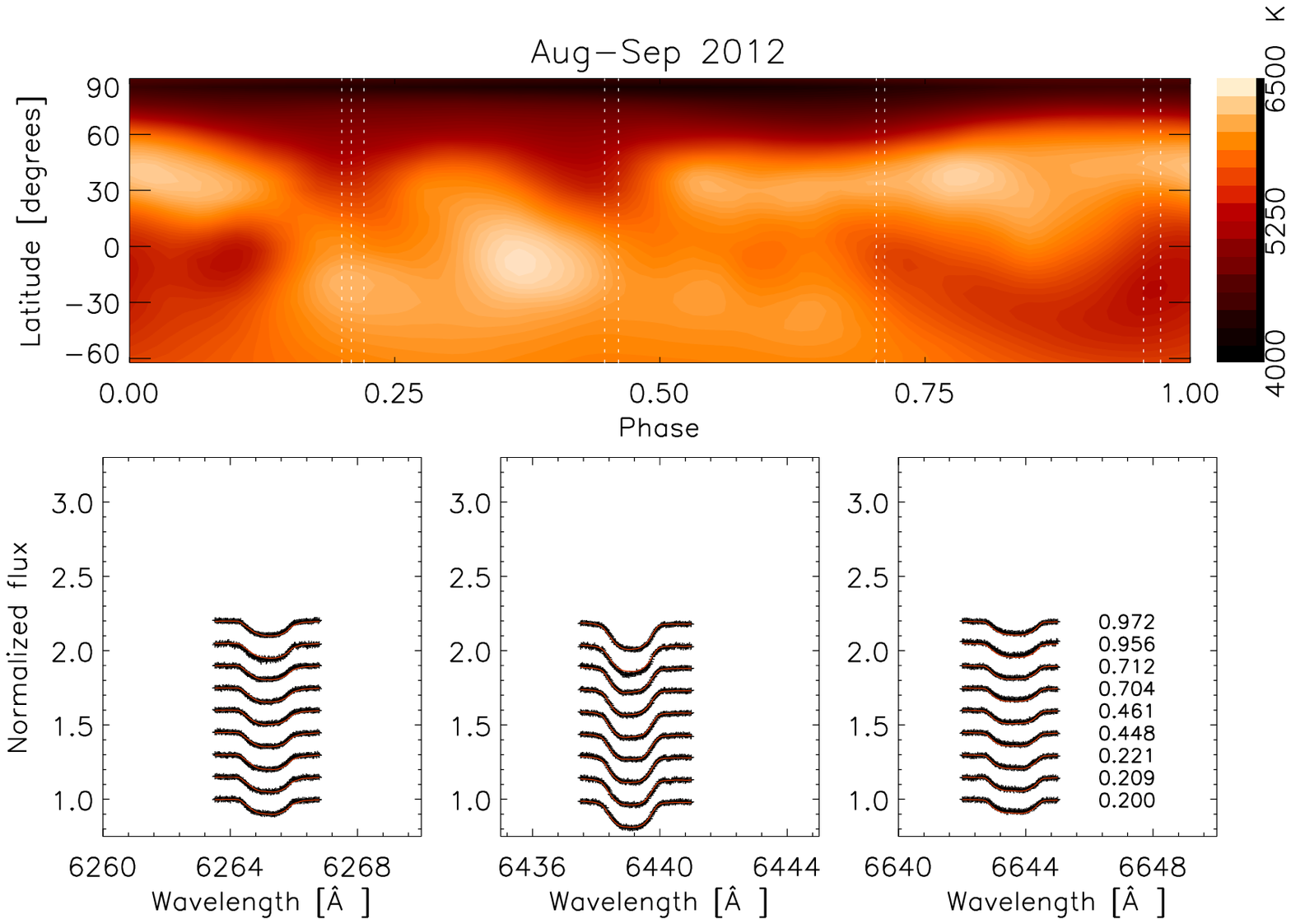} &
    \includegraphics[width=7cm]{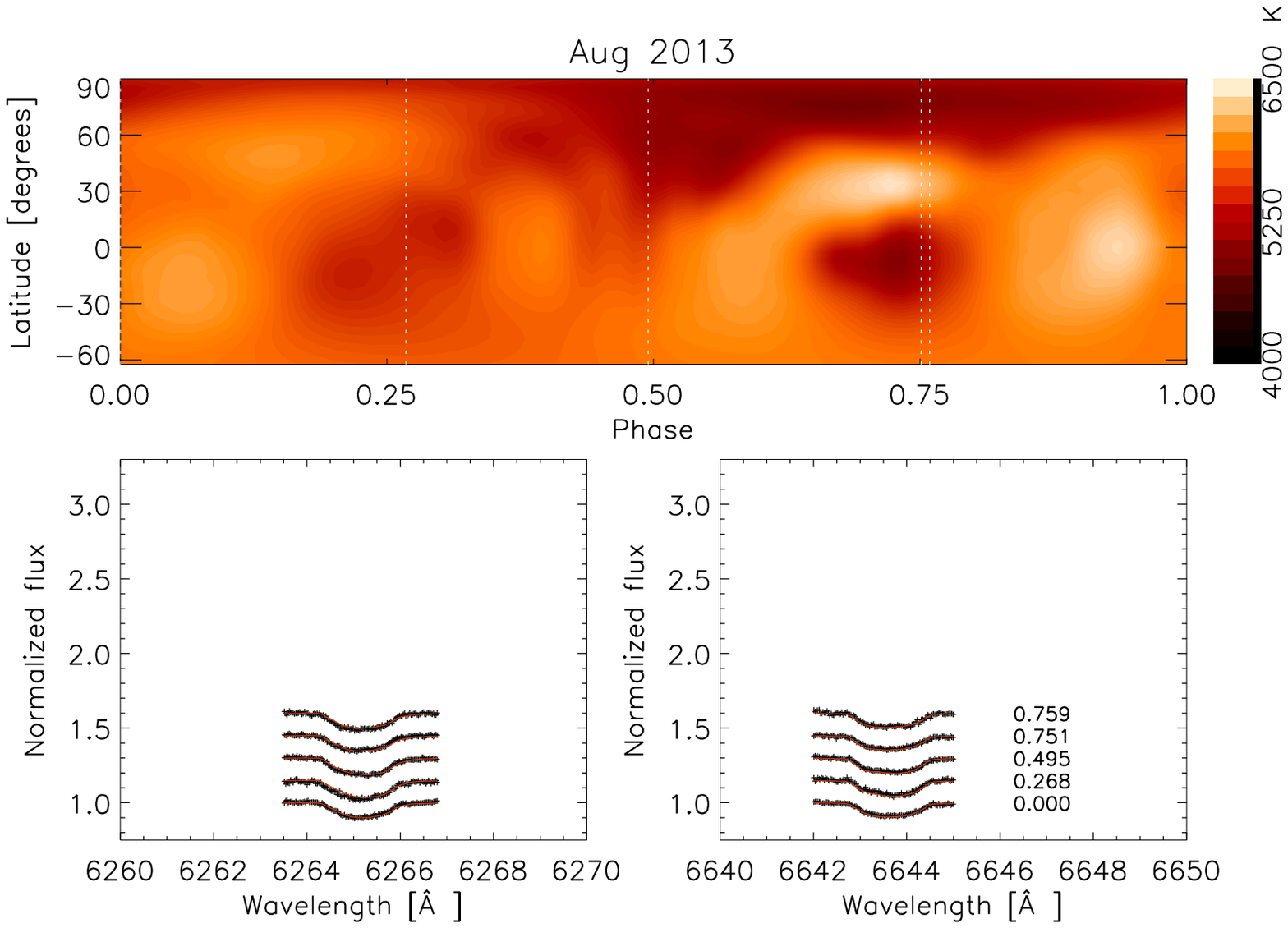} \\
    \vspace{-0.3cm}
    \includegraphics[width=7cm]{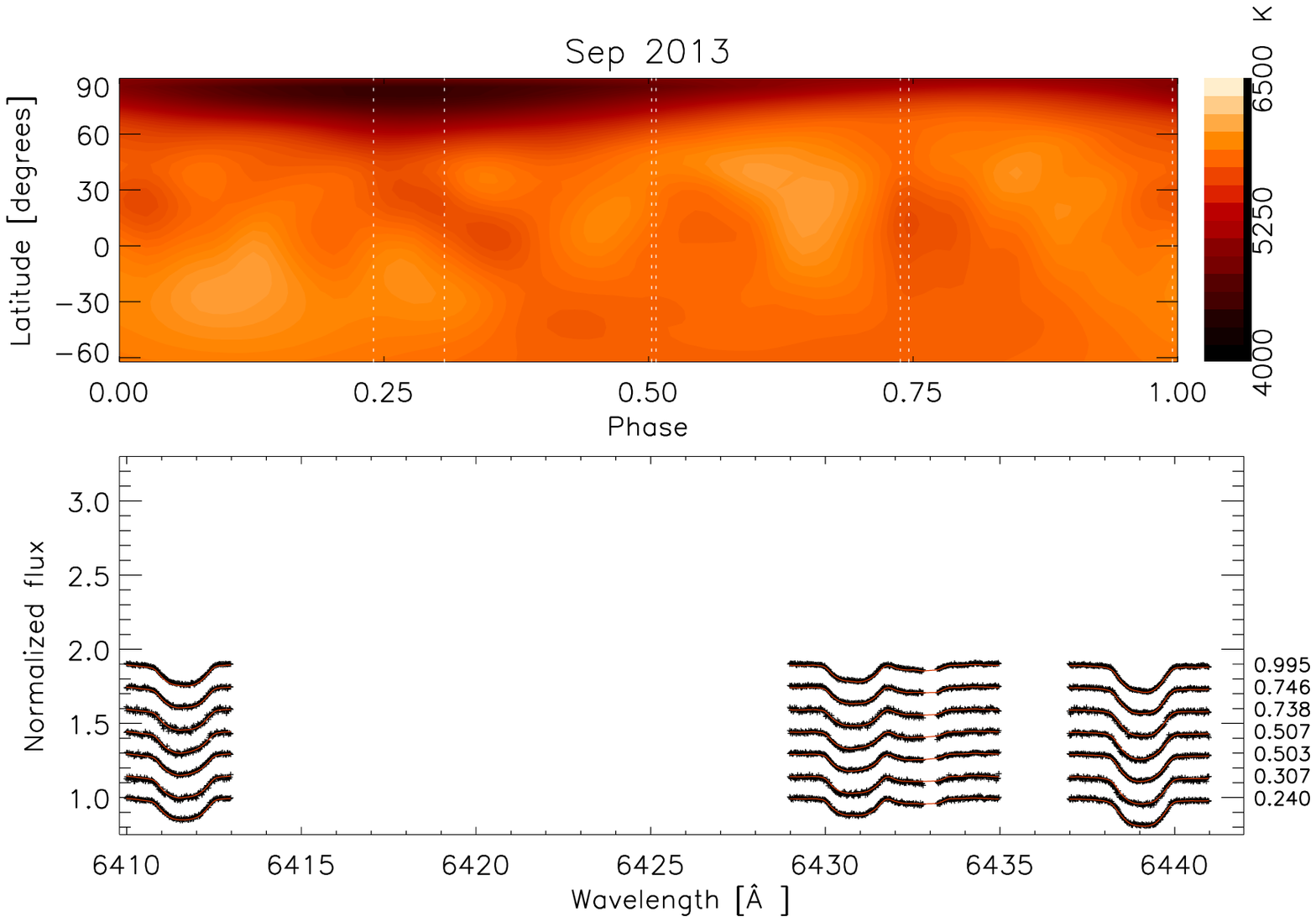} &
    \includegraphics[width=7cm]{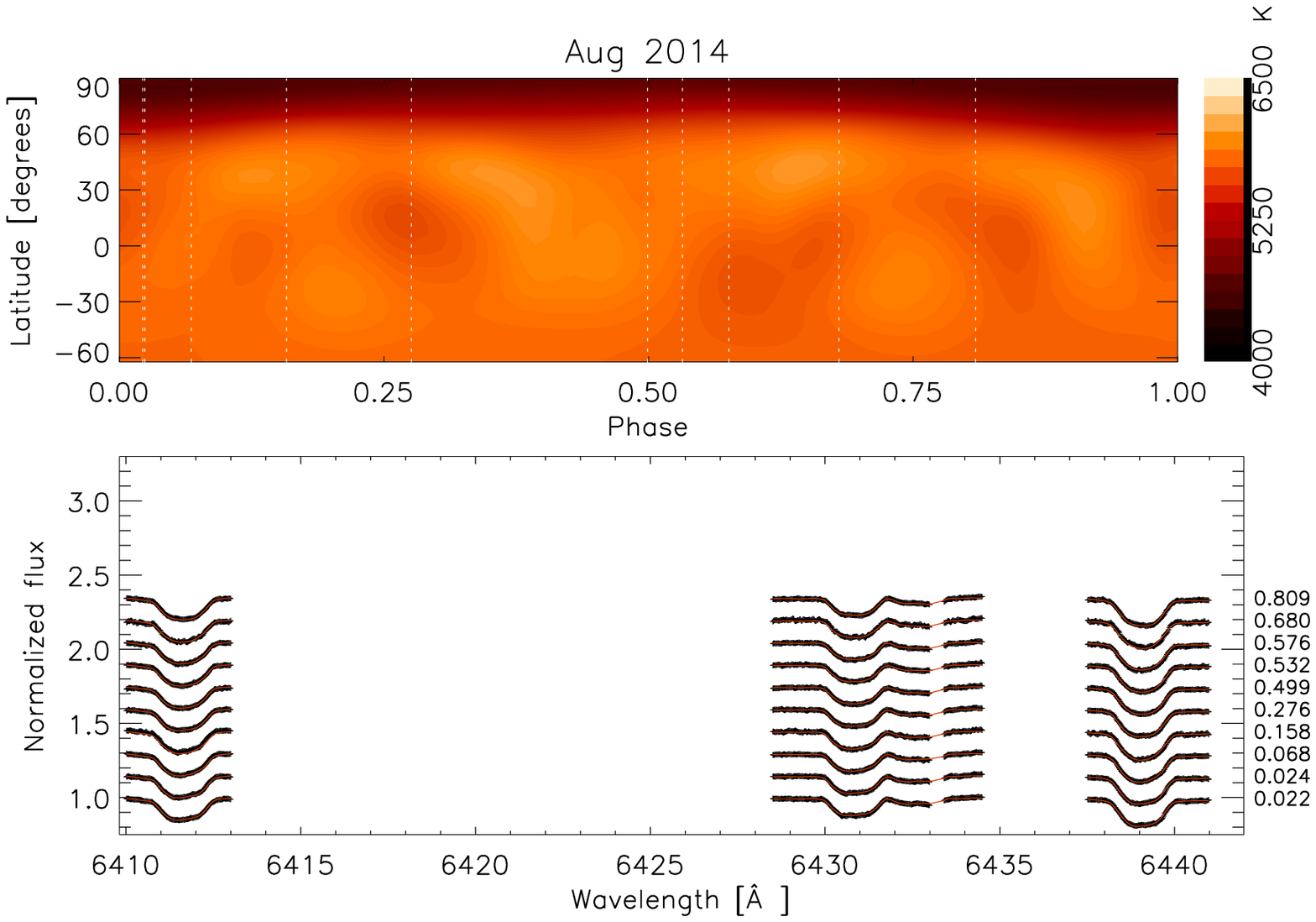} \\
    \vspace{-0.3cm}
    \includegraphics[width=7cm]{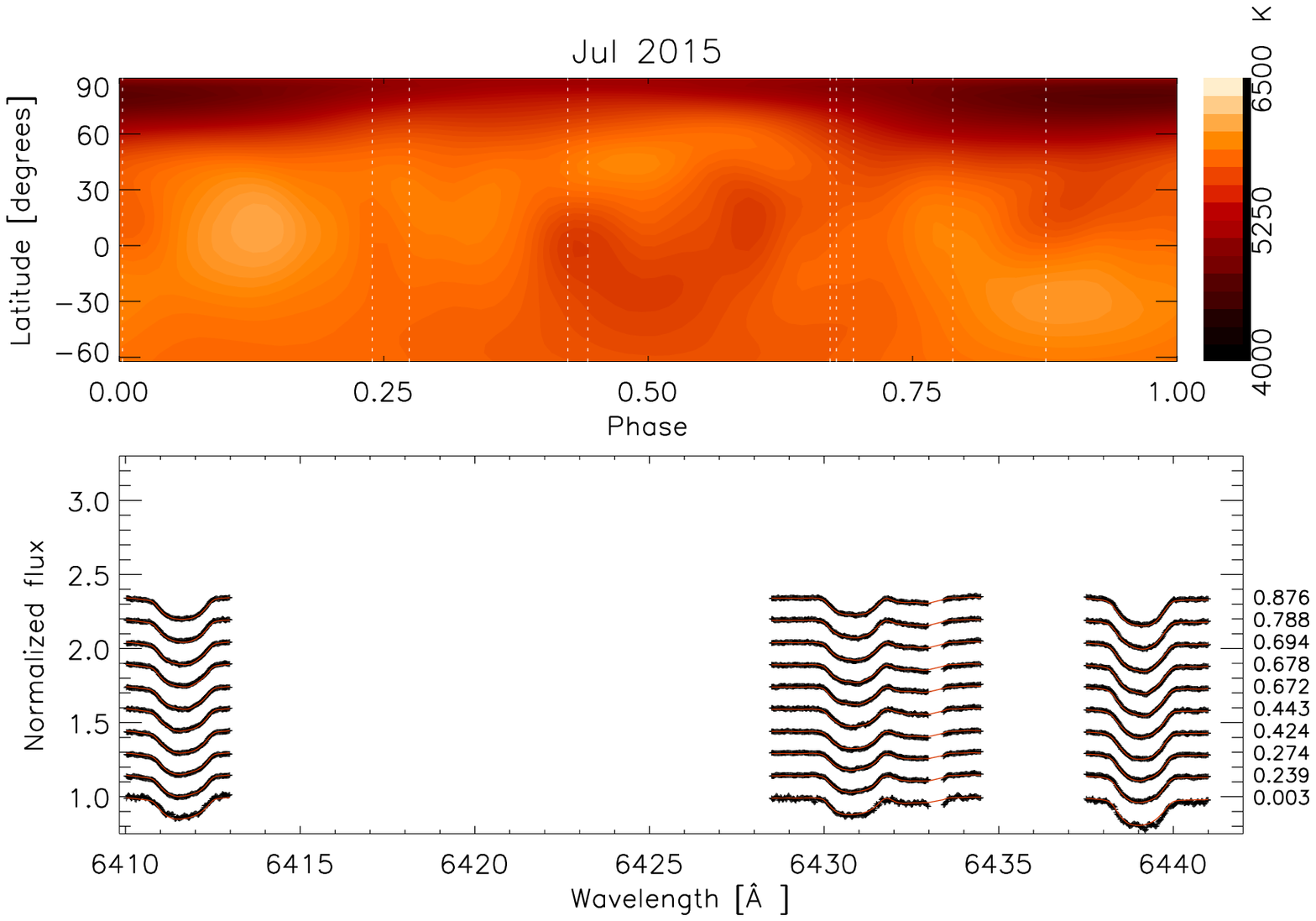} &
    \includegraphics[width=7cm]{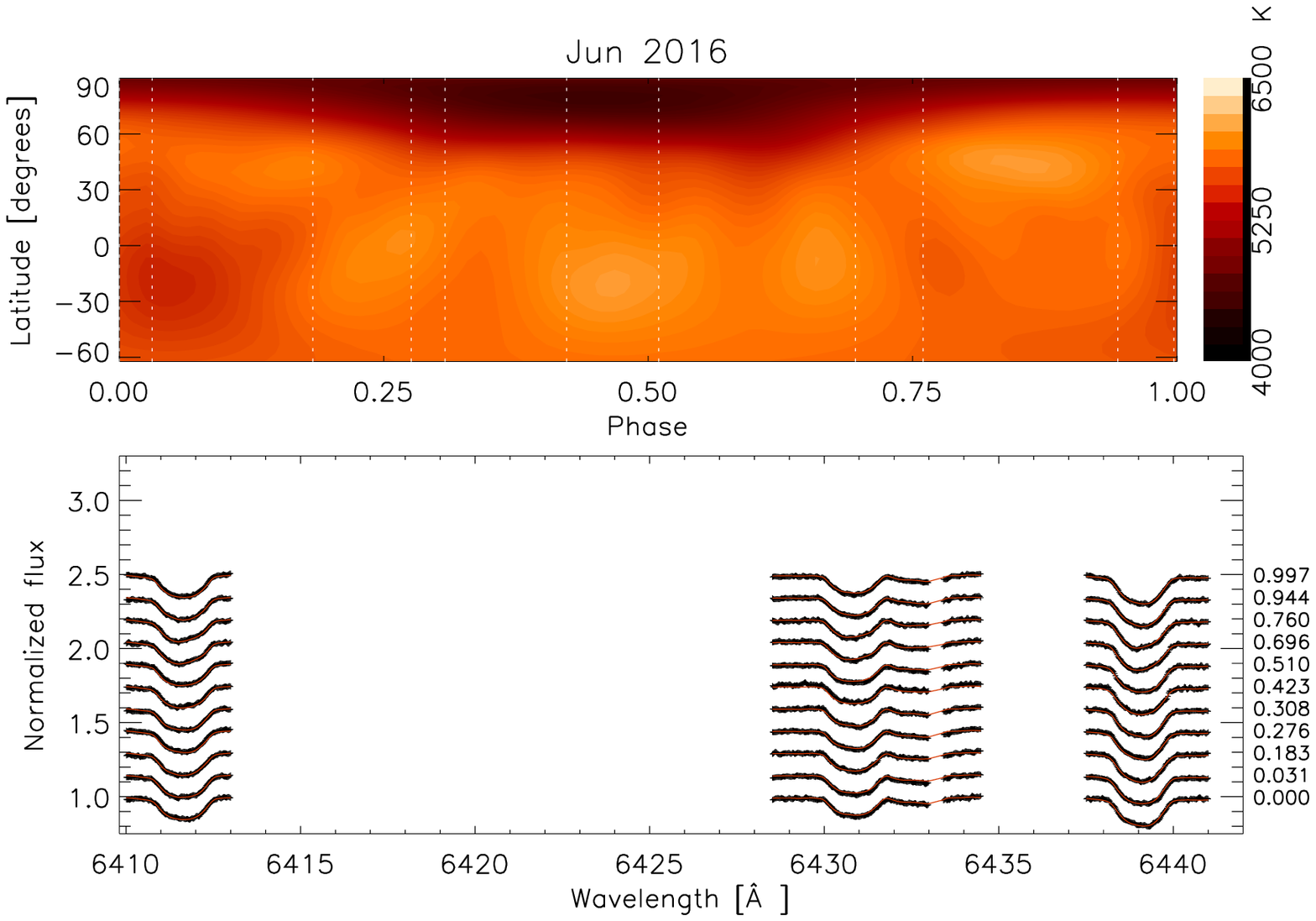} \\
    \vspace{-0.3cm}
    \includegraphics[width=7cm]{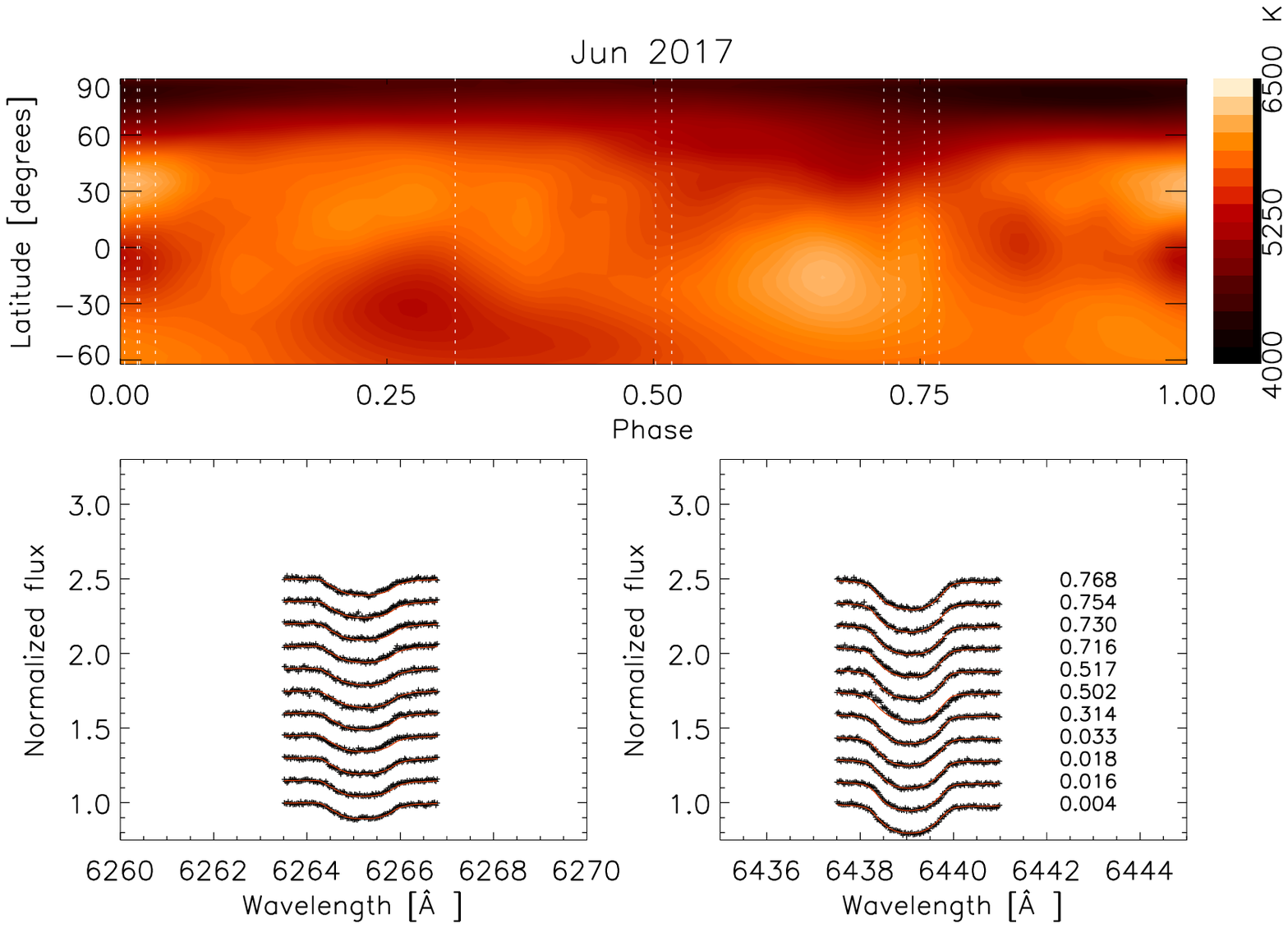} 
  \end{tabular}
  \caption{Same as Fig. \ref{DI_all}, but for 2009-2017.}
  \label{DI2}
\end{figure*}

\section{Results}

\subsection{Spottedness}

The surface temperature maps of V889 Her are presented in Figs. \ref{DI_all} and \ref{DI2}. From these maps, the spot filling factor $f_S$, i.e. the fraction of stellar surface covered with spots, has been calculated. We have defined all surface elements with temperature $T \leq 5300$ K to be treated as spots. This chosen threshold value is 
certainly 
higher than the typical temperature of sunspots, but lower than the temperature of the unspotted surface of V889 Her (in our Doppler maps the mean temperature is around 5750 K). The resulting filling factors are shown in Table \ref{f-factors} and Fig. \ref{Mfs}. It should be kept in mind, however,  that the choice of the limiting temperature for a spot is arbitrary, and the resulting filling factors 
will vary greatly with 
small changes in 
this threshold. 
A single 
data point may thus not be very reliable here; hence, we need a longer time series where systematic trends can be 
seen, 
if we want to follow the evolution of the spottedness. Furthermore, the filling factor may be influenced by possible hot features.

A stable, large high-latitude spot, centred in most cases a bit off the pole, is seen in all Doppler maps. The ones with very few rotational phases observed (e.g. September and October 1999) or those where the phase coverage is poor because the observed phases are close to each other (e.g. July-August 1999) cannot be regarded as reliable as those with better phase coverage (see Table \ref{obs} for the phase coverage for each set). 
For instance in 
the maps from 1999 in Fig. \ref{DI_all}, the surface features such as bright areas are clearly distributed around the observed phases in the longitudinal direction, 
indicating that they are artefacts typical for insufficient phase coverage. 
It is also notable, that with better phase coverage, the maps will in general be smoother. Thus, most of the small-scale variations in the maps with poorer phase coverage are probably also artefacts. The polar spot, however, is present in every map, and thus is most probably a real feature. It has also been observed in previous Doppler maps published of V889 
Her \citep[e.g.][]{strassmeier03,jarvinen08,frasca10}.
The polar spot seems to be a stable feature, lasting for decades but evolving in size.

\subsubsection{Comparison with photometry}

In the photometry, the clearly cyclic pattern has changed to a monotonic decreasing trend for the time between about 2007 to 2017. The decreasing trend was noted by \cite{lehtinen16}, but the trend  continued even more clearly in the new data analysed here until the brightness reached its minimum and turned to a rapid increase around 2017. The star seems thus to have reached a state of higher activity, possibly something similar to the Modern Maximum, the period from about 1950 to 2000 when the Sun was particularly active. It will be interesting to see if the star will remain in this state or return to its earlier activity state.

To compare photometric and spectroscopic results, both the filling factors $f_S$ and mean magnitudes $M$ are plotted in Fig. \ref{Mfs}. The hypothesis is that with decreasing brightness the spottedness should  increase since the luminosity of rapidly rotating young stars should be dominated by dark spots \citep{radick90}; however, this relation is  not seen very clearly. In the next section some possible reasons for this are discussed.

\subsubsection{Reliability of $f_S$} \label{reliability}

Inconsistencies in the spot filling factors may be caused by  
the change in spectral lines used. The level of $f_s$ changes between 2007 and 2008 from around 0.05 to 0.1. This could be explained by the different spectral lines rather than any real changes in the 
star since the preferred lines (3-6 in Table \ref{lin}) could not be used in the SOFIN sets from 2008 onward. However, 
in the
 data from 2004 and 2006 the Fe I and Fe II lines between 6411 and 6433 Å (lines 3-5) were not available, and still $f_S$ is on a  level similar to those in the other data sets prior to 
2008.

\begin{figure}
     \includegraphics[width=9.0cm]{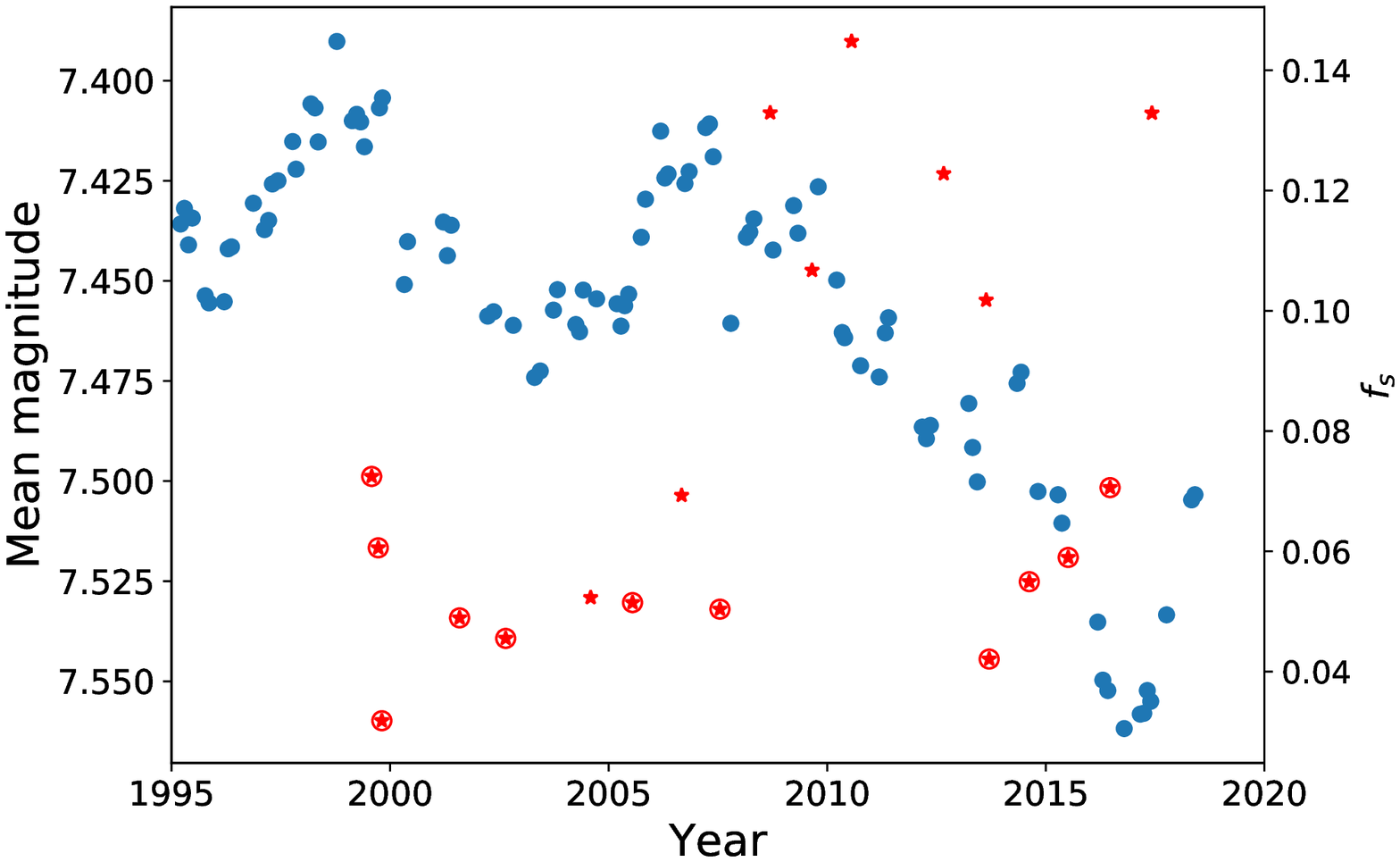} 
      \caption{Mean magnitudes of each independent set in the CPS analysis (blue dots) and spot filling factors (red stars). The sets where the preferred lines at 6411-6439 Å were used are highlighted with a circle.
              }
         \label{Mfs}
   \end{figure}

The FIES and HARPS observations between 2013 and 2016, where the full visible spectrum was available, yield    $f_S$ values comparable to those from  the sets before 2008. The preferred lines 3-6 have been used here, and the results differ notably from the SOFIN observations between 2008 and 2017, where different lines had to be used. In Fig. \ref{Mfs}, all SOFIN sets where lines 3-6 could be used are highlighted, and most of the other SOFIN sets clearly differ from these.

\begin{figure}
      \includegraphics[width=9.0cm]{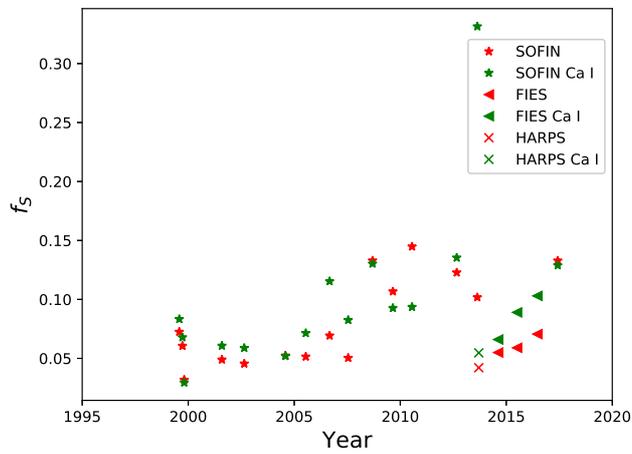} 
      \caption{Spot filling factors calculated with combined lines, compared to those where only the Ca I line was used. SOFIN, FIES, and HARPS observations are distinguished from each other.
              }
         \label{SOFIN_FIES}
     \end{figure}

As a test of the effect of using different lines in different data sets, we repeated the temperature inversion for all maps using only the Ca I line at 6439.0750 Å (line 6), which is the only line present in all the data sets. It is worth mentioning that as this is the strongest absorption line, it is formed highest up in the photosphere and may be affected by non-LTE effects more than the other lines, which makes it more difficult to model. 
Using only a single line causes more noise, but should reduce systematic differences related to different spectral regions. In Fig. \ref{SOFIN_FIES} we compare the filling factors from Fig. \ref{Mfs} to those where only the Ca I line was used. The data from the different instruments are shown separately.

Using only the Ca I line, the rise in the filling factors around 2008 seems smoother. 
There is also more variance in the data points produced with only Ca I, as is expected. 
In the data point from Aug 2013, $f_S$ exceeds 0.3, which is a clear outlier. In this particular set, we excluded the Ca I line from our combined line list, due to obvious instrumental errors.

The fact that no consistent trend is found from the filling factors could be due to the effects of using different 
line combinations, 
which yields a higher noise than the true changes in the spottedness. In the three FIES observations, which should be comparable to each other, however, a clear rising trend of $f_S$ is seen. The HARPS data from Sep 2013 also agree with this trend. 
The SOFIN observations, unfortunately, do not seem to be directly comparable  as they clearly show a  higher spottedness.

The differences in $f_S$ in Fig. \ref{Mfs} could be explained solely from the use of different lines, but in Fig. \ref{SOFIN_FIES} there is a notable difference between the SOFIN observations during and before 2013 and the following HARPS and FIES observations, even when only the Ca I line was used. In principle there should be no differences due to instrumentation.

We calculated a Pearson correlation coefficient for the mean magnitudes and $f_S$ values retrieved from the combined lines as $r = 0.18$, which implies a weak correlation. When using only those $f_S$ values where the preferred lines 3-6 could be used
, the correlation is much stronger; $r = 0.48$. This indicates that the choice of lines may significantly affect the results. The correlation coefficient between mean magnitudes and $f_S$ derived only with the Ca I line is $r = 0.24$.

To test for the effect of the choice of limiting temperature for a spot, we also calculated $f_S$ by using the limiting temperature 5000 K and 4700 K (instead of 5300 K) for a spot. This resulted in $r = 0.06$ for 5000 K and $r = 0.20$ for 4700 K. These differences (as $r = 0.18$ with 5300 K) further imply that the values of $f_S$ may not be that reliable.

To calculate the correlation coefficient, we linearly interpolated the mean magnitude of the epoch of the spectroscopic data points from the two closest phtometric data points.
Although $f_S$ does not show a very clear correlation to photometry in this case, we point out that there are other stars for which the correlation seems to be much clearer \citep[e.g.][]{hackman2012,hackman2018}.

\subsection{Mean temperature} 

\begin{figure}
      \includegraphics[width=9.0cm]{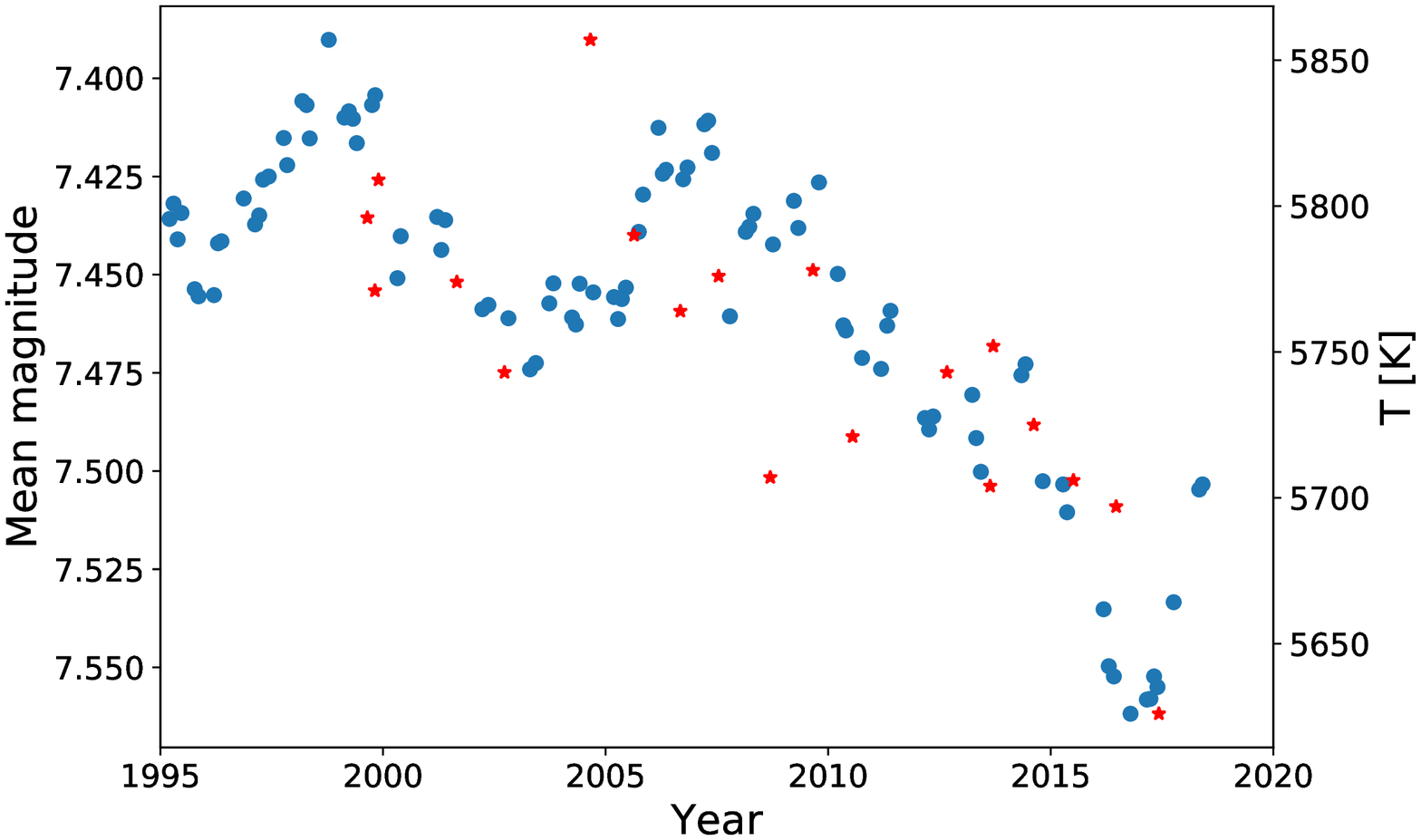} 
      \caption{Mean magnitudes (blue dots) plotted against the mean temperature of each spectral set (red stars).
              }
         \label{mtemp}
     \end{figure}

We also calculated the mean temperature of each Doppler map. Large spots should have a cooling effect on the temperature, and thus also lower the luminosity of the star. The photometric mean magnitudes are plotted against the spectroscopic mean temperatures in Fig. \ref{mtemp}. A much clearer correlation is seen here, suggesting that the decreasing luminosity of V889 Her is indeed caused by decreasing surface temperature, most likely in the form of increasing spottedness, although this is not clearly seen from the spot filling factors. The Doppler imaging method 
is 
very sensitive to changes in the mean temperature 
because this will affect the strength of the disk integrated line. 

We calculated a Pearson correlation coefficient for the mean magnitude and mean temperature values as $r = -0.60$. This is a significantly stronger correlation than between mean magnitudes and $f_S$. 
Thus, the mean temperature seems to be a more reliable indicator of the evolution of photospheric activity than the spot coverage, at least in this case. This is important since the mean temperature is independent of spot models or definitions of spotted regions, and is affected even by small features, which go easily undetected in the Doppler images and spot filling factors. Furthermore, the mean temperature is much less sensitive to Doppler imaging artefacts since they tend to be alternating cool and hot features which will eliminate each other in the mean temperature.

\subsection{Distribution of spots}

\begin{figure}
      \includegraphics[width=9cm]{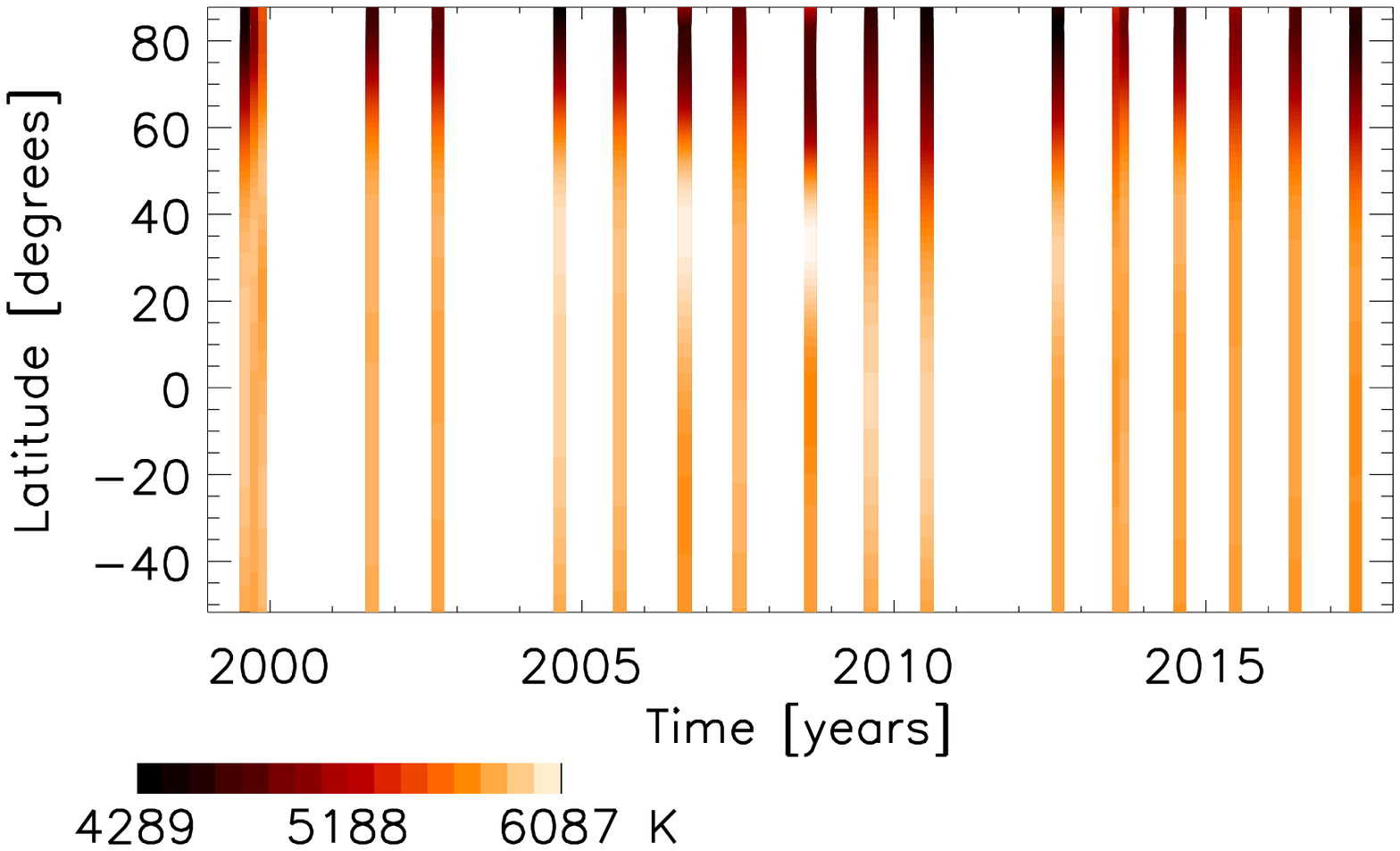}
      \caption{Surface temperature distribution of V889 Her from our 
Doppler images
averaged over all longitudes 
for 
each latitude. The time intervals of the observations are extended for clearer plots.
              }
         \label{lat}
   \end{figure}

\begin{figure}
      \includegraphics[width=9cm]{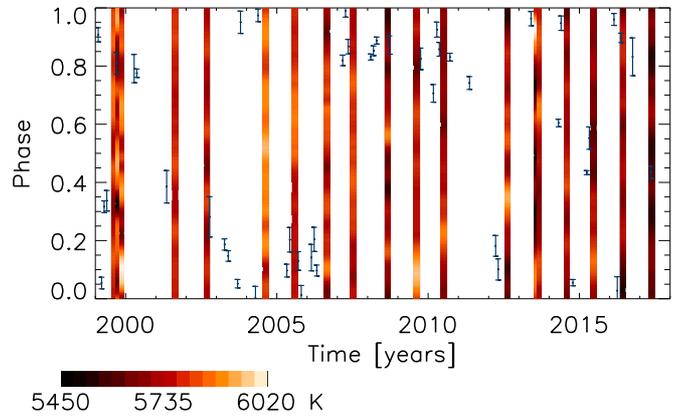} 
      \caption{Surface temperature distribution of V889 Her from our 
Doppler images
averaged over all latitudes 
for 
each longitude. The time intervals of the observations are extended for clearer plots. Phases of the photometric minima are also shown with their error bars.
              }
         \label{lon}
   \end{figure}

In Figs. \ref{lat} and \ref{lon} we show the averaged  surface temperature distribution of V889 Her over each latitude and longitude, respectively, to study changes in the average spot locations. The 
latitudinal 
spot distribution could reveal effects resembling the butterfly diagram of the Sun, whereas the 
longitudinal 
distribution could reveal active longitudes. However, no clear trends are revealed from our data. In Fig. \ref{lat} the mean spot latitude remains fairly constant.

In Fig. \ref{lon}, there  is some evolution in the 
longitudinal 
spot distribution, but no clear trends. It should also be kept in mind that possible differential rotation, which is not taken into account here, can affect the apparent longitudes of spots.

Figure \ref{lon} also shows the phases of the photometric minima, which should coincide with the dominating spot regions. This can be used as an independent check of the reliability of the spot longitudes. There is certainly not a perfect match between them, but in many cases they are located close to each other, suggesting that the spot longitudes are still fairly reliable. Photometric minima are also concentrated in most cases close to phase 0, although there is a great deal of  scatter. \cite{jetsu2018} showed that phases of photometric minima are accurate only if there is one dominating spot region.

\begin{table}
\centering
\caption{Filling factors $f_S$ and mean temperatures $T_{\mathrm{mean}}$ for each set of spectral observations. The lines used in each set are numbered as in Table \ref{lin}.}
\label{f-factors}
\begin{tabular}{c c c c}
\hline\hline
Set & Lines & $f_S$ & $T_{\mathrm{mean}}$ [K] \\
\hline
Jul -- Aug 1999 & 3,4,5,6 & 0.0725 & 5796\\ 
Sep 1999 & 3,4,5,6 & 0.0606 & 5771\\
Oct 1999 & 3,4,5,6 & 0.0318 & 5809\\
Jul -- Aug 2001 & 3,4,5,6 & 0.0489 & 5774\\
Aug 2002 & 3,4,5,6 & 0.0455 & 5743\\
Jul -- Aug 2004 & 1,2,6 & 0.0523 & 5857\\
Jul 2005 & 3,4,5,6 & 0.0515 & 5790\\
Aug -- Sep 2006 & 1,2,6 & 0.0693 & 5764\\
Jul 2007 & 3,4,5,6 & 0.0504 & 5776\\
Sep 2008 & 1,2,6,7 & 0.1329 & 5707\\ 
Aug -- Sep 2009 & 1,2,6,7 & 0.1067 & 5778\\
Jul 2010 & 1,2,6,7 & 0.1448 & 5721\\ 
Aug -- Sep 2012 & 1,2,6,7 & 0.1228 & 5743\\
Aug 2013 & 1,2,7 & 0.1018 & 5704\\
Sep 2013 & 3,4,5,6 & 0.0421 & 5752\\
Aug 2014 & 3,4,5,6 & 0.0550 & 5725\\
Jul 2015 & 3,4,5,6 & 0.0590 & 5706\\
Jun 2016 & 3,4,5,6 & 0.0706 & 5697\\
Jun 2017 & 1,2,6 & 0.1329 & 5626\\
\hline
\end{tabular}
\end{table}

\subsection{Chromospheric activity}

\noindent As an additional check for the activity level of V889 Her, we studied the evolution of the H$\alpha$ line at 6562.8 Å. The line is seen as an absorption line, but magnetic activity will increase H$\alpha$ emission from the chromosphere, which will weaken the strength of the absorption line. None of the SOFIN spectra, unfortunately, contain the line nor any of the other significant chromospheric lines, such as the H$\beta$ or Ca II H\&K lines, but from our FIES and HARPS spectra we get four data points for the average strength of the line. The equivalent width (EW), characterising the strength of a spectral line, is defined as

\begin{equation}
EW = \int_{\lambda_1}^{\lambda_2} \big(1 - \frac{F_{\lambda}}{F_c}\big)\mathrm{d}\lambda \approx \sum_{\lambda = 6560\AA}^{6566\AA} \big(1 - \frac{F_{\lambda}}{F_c}\big)\Delta\lambda,
\end{equation}

\noindent where $F_{\lambda}$ is the intensity at wavelength $\lambda$; $F_c$ is the continuum intensity, which is here normalised to 1; and $\Delta\lambda$ is the interval between wavelength points. The line strength is integrated over the line, which we have defined as the region between $\lambda_1 = 6560$ Å and $\lambda_2 = 6566$ Å. We have calculated this for every spectrum in each set, and taken the average value as the indicator for the chromospheric activity of the set. The EW values are shown in Table \ref{halpha}.

As we have not been able to use SOFIN spectra to check for chromospheric emission, we have too few data points to draw any reliable conclusions. Still, the EW from the FIES and HARPS spectra behave as would be expected: as the brightness of the star decreases from 2013 to 2016, the EW also decreases. As photospheric absorption dominates the H$\alpha$ line, the decrease in EW suggests that chromospheric activity is increasing. This supports the conclusion that V889 Her has indeed entered a state of enhanced activity.
\\

\begin{table}
\centering
\caption{Equivalent widths of the H$\alpha$ line for the FIES and HARPS spectra.}
\label{halpha}
\begin{tabular}{c c}
\hline\hline
Set & EW [Å] \\
\hline
Sep 2013 & 0.657 \\ 
Aug 2014 & 0.640 \\
Jul 2015 & 0.574 \\
Jun 2016 & 0.538 \\
\hline
\end{tabular}
\end{table}


\section{Conclusions}

We make the following conclusions from our study:

   \begin{enumerate}
      \item 
The clear cyclic behaviour of V889 Her seems to have 
been replaced with an almost monotonic luminosity decrease between 2007 and 2017 when the star reached its activity maximum, after which its luminosity  increased rapidly, although it is still notably lower than before 2007. We interpret it to be in a state of enhanced activity, possibly a grand maximum.
      \item 
The correlation between brightness and mean temperature is much clearer than the anti-correlation between brightness and spottedness.
      \item 
The polar spot, also reported  in previous Doppler imaging studies of V889 Her, is a fairly stable feature. It does not show any major changes even with the decreasing brightness and mean temperature.

   \end{enumerate}

From our results, it seems clear that the choice of which spectral lines are used in the Doppler imaging inversion may affect the outcome of the temperature map. This may cause serious problems when using instruments for which the spectral region varies between different observing seasons. There may also be some effects that depend on the instrument.

Our results indicate, that V889 Her has entered a state of decreased brightness and mean 
temperature. The cyclic behaviour seems to have continued after the long luminosity decrease, as the brightness has again increased in the most recent photometry. 
The fact that the
spottedness did not follow the same monotonic evolution may indicate that bright surface features
also influence the mean magnitude. Bright spots in Doppler images based on insufficient data are,
of course, not reliable. However,  the maps based on high $S/N$ and good phase coverage also show regions with
surface temperatures slightly higher than  expected for the unspotted surface. 
These hotter
surface areas could be generated by magnetic activity, but could also be of non-magnetic origin. \cite{kapyla2011} showed that both cooler and hotter regions may arise from large-scale vortices generated by
rotating turbulent convection. With rotation rates comparable to those of V889 Her, hotter anticyclonic vortices would be preferred from this mechanism.

\begin{acknowledgements}
The work of T.W. was financed by the Emil Aaltonen Foundation. M.J.K. was partially supported by the Academy of Finland Centre of Excellence ReSoLVE 272157. The Automated Astronomy Program at Tennessee State University has been supported by NASA, NSF, TSU, and the State of Tennessee through its Centers of Excellence programme. O.K. acknowledges support by the Knut and Alice Wallenberg Foundation, the Swedish Research Council, and the Swedish National Space Board. This work has made use of the VALD database, operated at Uppsala University, the Institute of Astronomy RAS in Moscow, and the University of Vienna. The authors thank the anonymous referee for the good comments and suggestions which improved the manuscript.
\end{acknowledgements}

\bibliographystyle{aa}
\bibliography{lahteet}

\end{document}